\documentclass[11pt,a4paper]{article}
\pdfoutput=1

\usepackage{jcappub}            
\usepackage[english]{babel}
\usepackage[utf8]{inputenc}
\usepackage[usenames,dvipsnames]{xcolor}
\usepackage{enumerate}
\usepackage[normalem]{ulem}
\usepackage{subcaption}
\usepackage{arydshln}
\usepackage{pdflscape}
\usepackage{amsbsy}
\usepackage{amsfonts}
\usepackage{mciteplus}          

\mciteSetSublistMode{s}
\mciteSetSublistLabelBeginEnd{\par}{\relax}{\relax}
\mciteSetMidEndSepPunct
  {\ifmciteBstWouldAddEndPunct.\else\fi\space}
  {\ifmciteBstWouldAddEndPunct.\else\fi}
  {\relax}

\DeclareOldFontCommand{\rm}{\normalfont\rmfamily}{\mathrm}
\DeclareOldFontCommand{\sc}{\normalfont\rmfamily}{\mathsc}
\DeclareOldFontCommand{\bf}{\normalfont\rmfamily}{\mathbf}
\DeclareOldFontCommand{\tt}{\normalfont\rmfamily}{\mathtt}

\newcommand{\bea}{\begin{eqnarray}} \newcommand{\eea}{\end{eqnarray}}

\newcommand{\re}[1]{(\ref{#1})}

\renewcommand{\sec}[1]{section \ref{#1}}
\newcommand{\fig}[1]{figure \ref{#1}}
\newcommand{\tab}[1]{table \ref{#1}}

\renewcommand{\a}{\alpha}
\renewcommand{\b}{\beta}

\renewcommand{\l}{\lambda}

\newcommand{\LCDM}{$\Lambda$CDM\ }
\newcommand{\GN}{G_{\mathrm{N}}}

\newcommand{\rmd}{\mathrm{d}}

\newcommand{\ie}{i.e.\ }
\newcommand{\eg}{e.g.\ }

\newcommand{\adot}{\dot{a}}

\newcommand{\wHtot}{w_{H\mathrm{tot}}}
\newcommand{\wDtot}{w_{D\mathrm{tot}}}

\newcommand{\av}[1]{\langle{#1}\rangle}

\newcommand{\Omn}{\Omega_{\mathrm{m0}}}

\newcommand{\ml}{\mathrm{l}}
\newcommand{\ms}{\mathrm{s}}
\newcommand{\dls}{d_\mathrm{ls}}
\newcommand{\dl}{d_\mathrm{l}}
\newcommand{\ds}{d_\mathrm{s}}
\newcommand{\tz}{\tilde{z}}

\title{Backreaction and FRW consistency conditions}

\author[a]{Francesco Montanari}

\author[a,b]{and Syksy R\"{a}s\"{a}nen}

\affiliation[a]{University of Helsinki, Department of Physics
and Helsinki Institute of Physics \\
P.O. Box 64, FIN-00014 University of Helsinki, Finland}

\affiliation[b]{Kobe University, Department of Physics, Kobe 657-8501, Japan}

\emailAdd{francesco.montanari@helsinki.fi}
\emailAdd{syksy.rasanen@iki.fi}

\abstract{
If the FRW metric is a good approximation on large scales, then the distance and the expansion rate, as well different notions of distance, satisfy certain consistency conditions.
We fit the JLA SNIa distance data to determine the expected amplitude of the violation of these conditions if accelerated expansion is due to backreaction. Adding cosmic clock and BAO expansion rate data, we also model-independently determine the current observational limits on such violation.

We find that the predicted maximum backreaction amplitude $|k_H|\lesssim1$ (95\% C.I.) is of the same order as the current observational constraints $|k_H|\lesssim1$, the precise numbers depending on the adopted fitting method (polynomials or splines) and stellar population evolution model. We also find that constraints on the value of $H_0$ determined from expansion rate data are sensitive to the stellar evolution model. We forecast constraints from projected LSST+Euclid-like SNIa plus Euclid galaxy differential age data.  We find improvement by factor of 6 for the backreaction case and 3 for the model-independent case, probing an interesting region of possible signatures.}


\begin{document}

\begin{flushleft}
	\hfill		 HIP-2017-22/TH \\
	\hfill		 KOBE-COSMO-17-11
\end{flushleft}

\setcounter{tocdepth}{2}

\setcounter{secnumdepth}{3}

\selectfont
\maketitle
\tableofcontents

\section{Introduction} \label{sec:intro}

Observations of the cosmic microwave background (CMB) and large scale
structure show that the universe is statistically isotropic (with
small anomalies \cite{Copi:2010na, *Ade:2013nlj}), and they are consistent with the universe being spatially homogeneous \cite{Hogg:2004vw, Scrimgeour:2012wt, Nadathur:2013mva, Laurent:2016eqo, Ntelis:2017nrj} (see also \cite{Labini:2009zi, *Labini:2010aj, *Labini:2011tj, *Labini:2011dv}), in line with inflationary predictions.
(See \cite{Clarkson:2007pz, Maartens:2011yx, Bonnor1986, *Stoeger1987, *Heavens:2011mr, Clarkson:2010uz, *Clarkson:2012bg, *Clifton:2011sn, Rasanen:2013swa, Rasanen:2014mca} for tests of the FRW metric and statistical homogeneity and isotropy.) A fundamental assumption in cosmology is that statistical homogeneity and isotropy imply that average properties of the universe over large scales are well described by the exactly homogeneous and isotropic Friedmann--Robertson--Walker (FRW) model. However, this is not necessarily the case: the effect of departures from exact homogeneity and isotropy on the averages is called backreaction \cite{Shirokov:1962, Buchert:1995fz, Ellis:1984bqf, *Ellis:1987zz, Ellis:2005uz, Rasanen:2011ki, Buchert:2011sx}. It has been suggested that backreaction could explain late-time accelerated expansion \cite{Buchert:1999mc, Wetterich:2001kr, Schwarz:2002ba, Rasanen:2003fy, *Rasanen:2004sa, Kolb:2004am}.
In Newtonian gravity, the effect of inhomogeneity and anisotropy reduces to a boundary term \cite{Buchert:1995fz}. In general relativity this is not the case, but backreaction on the expansion rate is small if the metric is perturbatively close to the same FRW metric everywhere \cite{Rasanen:2011ki} (see \cite{Green:2010qy, *Green:2013yua, *Green:2014aga, *Buchert:2015iva, *Green:2015bma, *Ostrowski:2015pzb, *Green:2016cwo} for a related debate). It is not clear whether this holds in the real universe at late times. Backreaction effects have been started to be studied with relativistic cosmological simulations \cite{Adamek:2013wja, *Adamek:2014gva, *Adamek:2014xba, *Adamek:2015eda, *Adamek:2016zes}, including ones that are fully non-linear \cite{Yoo:2012jz, *Bentivegna:2012ei, *Bentivegna:2013ata, *Adamek:2015hqa, *Bentivegna:2016fls} and without any symmetries \cite{Giblin:2015vwq, *Mertens:2015ttp, *Bentivegna:2015flc, *Giblin:2016mjp, *Bentivegna:2016stg, *Macpherson:2016ict, *Giblin:2017juu}, but the magnitude of backreaction remains an open question.

In addition to changing the expansion rate, deviations from exact homogeneity and isotropy also affect light propagation. In particular, they modify the FRW relation between the average expansion rate $H$ and the angular diameter distance $D_A$. This is a signature that cannot be mimicked by any FRW model (although extra dimensions can lead to a similar feature \cite{Ferrer:2005hr, *Ferrer:2008fp, *Ferrer:2009pq}). The FRW $H$-$D_A$ relation can thus be used to test the FRW metric if the expansion rate and the distance are measured independently \cite{Clarkson:2007pz}. Similar tests can be made for the relation between the angular diameter distance and the parallax distance \cite{Mccrea:1935, Rasanen:2013swa}, and the distance sum rule \cite{Rasanen:2014mca}. If backreaction is small, violation of the consistency conditions will be small. If backreaction is significant (in particular, if it leads to accelerated expansion), the degree to which the consistency conditions are violated depends on how the average expansion rate is modified. It can be argued that in a statistically homogeneous and isotropic universe the change of the $H$-$D_A$ relation due to backreaction can be approximated by keeping the mapping between the redshift and the affine parameter the same in the Sachs optical equations (just substituting the average expansion rate for the FRW expansion rate) and replacing the source term with its spatial average \cite{Rasanen:2008be, Rasanen:2009uw, Bull:2012zx, Lavinto:2013exa}, although the issue requires more study. The source term depends on the matter content via the combination $\rho+p$, where $\rho$ and $p$ are the energy density and pressure, respectively. This is not directly affected by a cosmological constant, so if matter can be approximated as dust and backreaction exactly mimics a cosmological constant plus FRW spatial curvature, \ie the expansion rate agrees with the \LCDM model (which we take to include the possibility of non-zero spatial curvature), then the consistency conditions are not violated.
Although there is at the moment no reliable calculation of the effect of inhomogeneity and anisotropy on the average expansion rate, there does not seem to be any reason for it to closely mimic the effect of a cosmological constant. In \cite{Boehm:2013qqa}, the expected violation of the consistency condition was estimated with a toy model and a model-independent low redshift expansion to be $\sim0.1\ldots1$. The consistency conditions have been observationally tested in \cite{Shafieloo:2009hi, Mortsell:2011yk, Sapone:2014nna, Rasanen:2014mca, Cai:2015pia, LHuillier:2016mtc, Yu:2016gmd, Li:2016wjm, Wei:2016xti}, with deviations of this order of magnitude allowed.

The better the distance-redshift relation agrees with the \LCDM model, the smaller is the allowed violation of the consistency conditions. Turning this around, distance data can be used to determine how large violations are still allowed. If the maximum amplitude were to fall below the amplitude expected from backreaction by general theoretical arguments, this would be strong evidence against backreaction even in the absence of a precise prediction for the expansion rate.

We do a model-independent fit to the JLA dataset of type Ia supernova (SN) \cite{Betoule:2014frx} to determine the luminosity distance $D_L$, find the corresponding expansion history using the backreaction relation of \cite{Rasanen:2008be} between $D_L$ and $H$ and calculate the predicted violation of the consistency condition. We also test the consistency condition using independent observations of $D_L$ and $H$ without assuming a theoretical model and compare to the prediction.

In \sec{sec:theory} we introduce the FRW consistency conditions and
the backreaction equations, in \sec{sec:data} we go over the data and in
\sec{sec:method} we discuss the fitting methods. In
\sec{sec:res} we give our results for the consistency condition both as predicted by the backreaction $H-D_A$ relation and as determined solely from observations, consider the value of $H_0$ implied by $H(z)$ data and do a forecast for next generation data. In \sec{sec:conc} we summarise our findings.

\section{Theory} \label{sec:theory}

\subsection{FRW consistency conditions} \label{sec:cons}

If the universe (more precisely, light propagation on scales larger than the cosmological homogeneity scale) is described by the FRW metric and the geometrical optics approximation holds, certain consistency conditions relate the expansion rate and distance (as well as different notions of distance) to each other. These relations are purely geometrical, they are independent of the matter content and the dynamical relation between matter and spacetime (\ie the Einstein equation).

The consistency conditions can be expressed in terms of the spatial curvature parameter $k$ that is constant in the FRW case. One condition relates the dimensionless expansion rate $h(z)\equiv H(z)/H_0$ (where $H(z)$ is the Hubble parameter as a function of redshift $z$; the subscript $0$ denotes present value throughout) and the dimensionless comoving angular diameter distance $d(z)\equiv (1+z) d_A(z)\equiv (1+z) H_0 D_A(z)$ \cite{Clarkson:2007pz}:
\bea \label{kH}
  k_H(z) &\equiv& \frac{1 - h^2 d'^2}{d^2} \ ,
\eea
where prime denotes derivative with respect to the redshift $z$ and the subscript $H$ indicates that the relation involves the Hubble rate.

Another consistency condition relates the angular diameter distance and the parallax distance $d_P$ \cite{Rasanen:2013swa}
\bea \label{kP}
  k_P(z) &\equiv& \frac{1}{d^2} - \left( \frac{1}{d_P} - 1 \right)^2 \ ,
\eea
where the subscript $P$ indicates that the relation involves the parallax distance.

A third condition can be derived from the sum rule between distances. In the spatially flat FRW case, the comoving angular diameter distance $\ds\equiv d(z_\ms)$ from $z=0$ to $z=z_\ms$ is simply the sum of the distance $\dl\equiv d(z_\ml)$ from $z=0$ to $z_\ml$ and the distance $\dls\equiv d(z_\ml,z_\ms)$ from $z_\ml$ to $z_\ms$. In the spatially curved FRW case, the sum rule is more complicated, and the following combination is constant \cite{Rasanen:2014mca}
\bea \label{kS}
  k_S(z_\ml,z_\ms) &\equiv& - \frac{\dl^4 + \ds^4 + \dls^4 - 2 \dl^2 \ds^2 - 2 \dl^2 \dls^2 - 2 \ds^2 \dls^2}{4 \dl^2 \ds^2 \dls^2} \ ,
\eea
where the subscript $S$ indicates that the relation comes from the distance sum rule.

In a FRW universe, the three functions $k_H$, $k_P$ and $k_S$ are constant and equal to minus the spatial curvature density parameter today, $-\Omega_{K0}$. If the spacetime is not well described by a FRW metric, they in general vary with $z$ and are different from each other. If it were observed that any of them depend on $z$, or that any two of them are not equal, this would indicate that the FRW metric approximation is not valid. Observational constraints on $k_H$ have been reported in \cite{Shafieloo:2009hi, Mortsell:2011yk, Sapone:2014nna, Rasanen:2014mca, LHuillier:2016mtc, Yu:2016gmd, Li:2016wjm, Wei:2016xti} and on $k_S$ in \cite{Rasanen:2014mca}. There are currently no observations of the parallax distance over cosmological distances, and thus no constraints on $k_P$, but this is expected to change with upcoming data from the Gaia satellite\footnote{\url{http://sci.esa.int/gaia/}} \cite{Rasanen:2013swa}.

\subsection{Backreaction}

\subsubsection{Redshift and expansion rate}

Let us now introduce the relation between the distance and the expansion rate in the backreaction case. If matter can be approximated as dust and evolution of structures is slow compared to the homogeneity scale, it can be argued \cite{Rasanen:2008be, Rasanen:2009uw, Buchert:2011sx} (see also \cite{Rasanen:2011bm, Bull:2012zx, Lavinto:2013exa, Koksbang:2017arw}) that, analogously to the FRW case, the redshift is given by
\bea \label{z}
  1 + z &=& a^{-1} \ ,
\eea
where $a(t)$ is the scale factor defined so that the volume of the hypersurface of statistical homogeneity and isotropy is proportional to $a(t)^3$, where $t$ is the time that is constant on the hypersurface. This relation depends on the cancellation of expansion rate fluctuations and matter shear along the null geodesic. The average expansion rate is $H=\adot/a$, where dot denotes derivative with respect to $t$.

The angular diameter distance can be solved from the Sachs optical equations. Assuming that null shear can be neglected and the Einstein equation holds, we have
\bea \label{Sachs}
  \frac{\rmd^2 d_A}{\rmd\l^2} = - 4\pi\GN ( \rho + p ) E^2 d_A \ ,
\eea
where $\l$ is the affine parameter, $\GN$ is Newton's constant, $\rho$ and $p$ are the energy density and pressure, respectively, and $E$ is photon energy. We normalise the affine parameter as $\l\rightarrow E_\mathrm{o}^{-1}\l$, where $E_\mathrm{o}$ is photon energy at the observer. The dimensionless luminosity distance is $d_L\equiv H_0 D_L=(1+z)^2d_A$ \cite{Etherington:1933, Ellis:1971pg}.

In the backreaction case, it can be argued that the average expansion rate gives the relation between the affine parameter $\l$ and the redshift in the same way as in the FRW case, with the average expansion rate in place of the FRW expansion rate, $\rmd\l=-(1+z)^{-2} H(z)^{-1} \rmd z$, and the source is given by the spatial average, so \re{Sachs} reduces to \cite{Rasanen:2008be, Rasanen:2009uw}
\bea \label{Sachsbr}
  h \frac{\rmd}{\rmd z} \left[ (1+z)^2 h d'_A \right] &=& - \frac{3}{2} \Omn (1+z)^3 d_A \ ,
\eea
where we have assumed that the matter can be approximated as dust ($p=0$), and $\Omn\equiv8\pi\GN\av{\rho}_0/(3 H_0^2)$, where $\av{}$ stands for spatial average. The initial conditions are $d_A(0)=0$, $d_A'(0)=1$. The relation \re{Sachsbr} differs from the FRW case with general matter content in that there is no pressure term on the right-hand side. Thus, if backreaction were to mimic the contribution of a cosmological constant plus FRW spatial curvature so that the expansion rate is given by $h^2 = \Omn (1+z)^{3} + \Omega_{K0} (1+z)^2 + 1 - \Omn - \Omega_{K0}$, the distance would be the same as in the FRW case, and $k_H$ defined in \re{kH} would have the constant value $k_H(z)=-\Omega_{K0}$. Conversely, if backreaction changes the expansion rate in a different way, the distance will be different from the FRW case (where not only $H(z)$ but also the right-hand side source term would change), and $k_H(z)$ will not be constant.

Solving \re{Sachsbr}, we get $h(z)$ in terms of $d_A(z)$,
\bea \label{hbr}
  h(z)^2 &=& \frac{1}{ (1+z)^4 (d'_A)^2 } \left[ 1 - 3 \Omn \int_0^z \rmd \tz ( 1 + \tz )^5 d_A(\tz) d'_A(\tz) \right] \ .
\eea
The FRW analogue of this equation is $h^2=(1+\Omega_{K0} d^2)/(d')^2$, solved from \re{kH}. (These two results agree if and only if the expansion rate has the \LCDM form given above.) In the FRW case, we need $d_A(z)$ and the value of $\Omega_{K0}$ to determine $h(z)$. At first sight, it might seem that in the backreaction case $\Omn$ has a similar role as $\Omega_{K0}$ in the FRW case. However, $\Omn$ is not a free parameter. In realistic cosmologies, $d_A$ has a maximum, so for $h$ to remain finite, the zero of $d_A'$ in the denominator must coincide with the zero of the numerator, which fixes $\Omn$.

We characterise the distance and the expansion rate with two effective equations of state. The effective expansion rate equation of state is defined as \cite{Boehm:2013qqa}
\bea \label{wHtot}
  \wHtot(z) &\equiv& \frac{2}{3} (1+z) \frac{h'}{h} - 1 \ ,
\eea
and the distance equation of state $\wDtot(z)$ is defined in the same way, but substituting $h=1/d'$. The function $\wHtot(z)$ is the total equation of state of matter in the spatially flat FRW model that has the expansion rate $h(z)$, and $\wDtot(z)$ is the equation of state corresponding to the spatially flat FRW model with distance $d(z)$. If the universe is well described by the spatially flat expansion FRW model, we have $\wHtot(z)=\wDtot(z)$, otherwise they will not in general agree. In the context of FRW models, it is more common to discuss the equation of state of a dark energy component alone rather than the total equation of state. However, if backreaction first slows down and then speeds up the expansion rate (or vice versa), the effective dark energy equation of state diverges at the transition. If backreaction explains the accelerated expansion, such evolution is expected \cite{Rasanen:2006zw, *Rasanen:2006kp, Boehm:2013qqa}. In fact, such a feature turns out to be common in model-independent fits to the real distance and expansion rate data, making it impossible to assign a finite effective dark energy equation of state, while the total equation of state is well-defined and finite.

\subsubsection{Consistency conditions in the backreaction case} \label{sec:brcon}

Given $d(z)$ from the data, we determine $h(z)$ from \re{hbr} and, using that together with $d(z)$, get $k_H(z)$ from \re{kH}. Let us now find the expressions for $k_S(z_\ml,z_\ms)$ and $k_P(z)$ in terms of $d(z)$ in the backreaction case. To express $k_S(z_\ml,z_\ms)$ in terms of $d(z)$, we use the fact that the angular diameter distance $d_A(z_\ml,z)=(1+z)^{-1}d(z_\ml,z)$ from $z_\ml$ to $z$ satisfies the Sachs equation \re{Sachs} for all values of $z_\ml$, which correspond to different initial conditions. In particular, $d_A(0,z)=d_A(z)$. As noted in \cite{Rosquist:1988}, any two solutions of the Sachs equation \re{Sachs} can be expressed (when null shear can be neglected) in terms of each other, and we can write
\bea \label{gf}
  d_A(\l_\ml,\l) = B(\l_\ml) d_A(\l) \int_\l^{\l_\ml} \frac{\rmd\tilde\l}{d_A(\tilde\l)^2} \ ,
\eea

\noindent where $B(\l_\ml)$ is an integration constant. Taking a derivative of \re{gf} and applying the initial condition $\frac{\rmd d_A(\l_\ml,\l)}{\rmd\l}|_{\l=\l_\ml}=-H_0 (1+z_\ml)$, we get $B(\l_\ml)= H_0 d_A(z_\ml)$, giving us $\dls$ in terms of $d(z)$,
\bea
  \label{dlsl} d(z_\ml,z_\ms) &=& H_0 d(z_\ml) d(z_\ms)  \int_{\l_\ms}^{\l_\ml} \frac{\rmd\l}{d_A(\l)^2} \\
  \label{dls} &=& d(z_\ml) d(z_\ms) \int_{z_\ml}^{z_\ms} \frac{\rmd z}{h(z) d(z)^2} \ ,
\eea
where on the second line we have used the relation $\rmd\l=-(1+z)^{-2} H(z)^{-1} \rmd z$.
Given observations of $d(z)$, we can find $h(z)$ from \re{hbr} and use these in \re{dls} to find $d(z_\ml,z_\ms)$.
Together with $d(z)$, we then get $k_S(z_\ml, z_\ms)$ from \re{kS}.

For the parallax distance, we have $\frac{\rmd D_P^{-1}}{\rmd\l} = \frac{1}{D_A^2}$ \cite{Rasanen:2013swa}. Integrating, we get
\bea
  \label{dpl} d_P(z)^{-1} &=& A + H_0^{-1} \int^\l \frac{\rmd\tilde\l}{D_A(\tilde\l)^2} \\
  \label{dp} &=& 1 - \int^z \frac{\rmd\tz}{h(\tz) d(\tz)^2} \ ,
\eea
where we have on the second line again used $\rmd\l=-(1+z)^{-2} H(z)^{-1} \rmd z$, and fixed the integration constant $A$ to unity by comparing \re{kP} and \re{hbr} around $z=0$ to next-to-leading order. The value of $A$ doesn't matter for us, though, because the indefinite integral is not suited for numerical evaluation. However, noting that \re{dpl} involves the same integral as \re{dlsl}, we can write
\bea \label{PandS}
  \frac{\dls}{\ds} &=& d(z_\ml) \left( \frac{1}{d_P(z_\ml)} - \frac{1}{d_P(z_\ms)} \right) \ ,
\eea
where $A$ drops out. Expressing $d_P$ in terms of $d$ and $k_P$ using \re{kP}, inserting into \re{PandS} and comparing to \re{kS} shows that $k_P(z)=k_S(z,z)$. It immediately follows that if $k_S$ is constant, $k_P$ is constant and equal to $k_S$. By using \re{kP}, \re{kS} and \re{PandS}, we can show that a constant $k_P$ also implies that $k_S$ is constant and equal to $k_P$. If we assume $\rmd\l=-(1+z)^{-2} H(z)^{-1} \rmd z$, these conditions are also equivalent to $k_H$ being constant and $k_H=k_P=k_S$.

\section{Data} \label{sec:data}

\subsection{Distance from type Ia supernovae}

We determine $d(z)$ from SNe Ia light-curve data of the SDSS-II/SNLS3 Joint Light-curve Analysis (JLA) \cite{Betoule:2014frx}. The catalogue includes light-curve parameters and their covariances for 740 SNe Ia in the redshift range $0.01<z<1.3$.  Another commonly used SNIa catalogue is Union2.1, which includes 560 SNe in the range $0.015<z<1.4$ \cite{Suzuki:2011hu}. The larger redshift range of Union2.1 would be an advantage in reconstructing the expansion rate (which requires determination of the zero of $d_A'$, which is typically at $z>1$) compared to JLA. However, in the Union2.1 catalogue the colour and stretch light-curve parameters are fixed through a fit to a reference spatially flat $\Lambda$CDM model, and the covariance matrices necessary to marginalise over the related coefficients are not provided. Hence, if the data are used to fit different cosmologies, there is an unquantified model-dependence (see \eg \cite{Nadathur:2010zm}). Analysis of mock data also suggests that the slightly larger redshift range would not make a significant difference for our results. We therefore use the JLA dataset only.

When determining the luminosity distance from the JLA data, we simultaneously fit cosmological parameters and the coefficients of the light-curve parameters. The JLA public dataset includes the light-curve parameters themselves (and their covariance) constrained using the SALT2 method \cite{Guy:2007dv} (which does not require an assumption about a cosmological model). The results depend on the modelling of the light-curves; for discussion of light-curve modelling and the impact of systematics on SNIa datafitting, see \cite{Bengochea:2010it, *Li:2010du, *March:2011xa, *Lago:2011pk, *Giostri:2012ek, Kessler:2012gn, Nielsen:2015pga, Shariff:2015yoa, Rubin:2016iqe, Dam:2017xqs, Tutusaus:2017ibk}. The mapping of light-curve variation to the physical processes that the light-curve fitters model is not unique \cite{Marriner:2011mf} and redshift-dependence of light-curve parameters can introduce degeneracy with cosmological parameters. Also, properties of SNIae depend on the host galaxies in a manner that is not completely understood and that has an impact on the inferred equation of state \cite{Betoule:2014frx, Campbell:2016zzh}. There may also be residual model-dependence from correcting for selection biases \cite{Dam:2017xqs}.

Given these caveats, the interpretation of the precise results of SNIa data analysis requires caution, as demonstrated by comparison of light-curve fitters \cite{Bengochea:2010it, *Li:2010du, *March:2011xa, *Lago:2011pk, *Giostri:2012ek} and datafitting methods \cite{Nielsen:2015pga, Shariff:2015yoa}. Nevertheless, such errors are not expected to change the qualitative picture and we are interested in the order of magnitude of the constraints rather than precise parameter estimates. Also, these errors are expected to be subdominant to the choice of how to determine $d(z)$ and $d'(z)$ from the data model-independently.

\subsection{Expansion rate from cosmic clocks and BAO} \label{sec:exp}

\begin{table}[t]
  \centering
  \begin{tabular}[t]{ccc}
    \hline
    \hline
    $z$ & $H(z)$ [km/s/Mpc] & Method \\
    \hline
    0.070  & $69   \pm 19.6$  & BC03 \cite{Zhang:2012mp}    \\
    0.120  & $68.6 \pm 26.2$  & BC03 \cite{Zhang:2012mp}    \\
    0.179  & $75  \pm 4 $     & BC03 \cite{Moresco:2012jh}  \\
    0.199  & $75  \pm 5 $     & BC03 \cite{Moresco:2012jh}  \\
    0.200  & $72.9 \pm 29.6$  & BC03 \cite{Zhang:2012mp}    \\
    0.280  & $88.8 \pm 36.6$  & BC03 \cite{Zhang:2012mp}    \\
    0.32   & $78.6 \pm 2.7$   & BAO \cite{Alam:2016hwk}     \\
    0.352  & $83  \pm 14$     & BC03 \cite{Moresco:2012jh}  \\
    0.3802 & $83.0 \pm 13.5$  & BC03 \cite{Moresco:2016mzx} \\
    0.4004 & $77.0 \pm 10.2$  & BC03 \cite{Moresco:2016mzx} \\
    0.4247 & $87.1 \pm 11.2$  & BC03 \cite{Moresco:2016mzx} \\
    0.4497 & $92.8 \pm 12.9$  & BC03 \cite{Moresco:2016mzx} \\
    0.4783 & $80.9 \pm 9.0 $  & BC03 \cite{Moresco:2016mzx} \\
    0.480  & $97 \pm 62$      & BC03 \cite{Stern:2009ep}    \\
    0.57   & $96.9 \pm 2.8$   & BAO \cite{Alam:2016hwk}     \\
    0.593  & $104 \pm 13$     & BC03 \cite{Moresco:2012jh}  \\
    0.680  & $92  \pm 8 $     & BC03 \cite{Moresco:2012jh}  \\
    0.781  & $105 \pm 12$     & BC03 \cite{Moresco:2012jh}  \\
    0.875  & $125 \pm 17$     & BC03 \cite{Moresco:2012jh}  \\
    0.880  & $90 \pm 40$      & BC03 \cite{Stern:2009ep}    \\
    1.037  & $154 \pm 20$     & BC03 \cite{Moresco:2012jh}  \\
    1.363  & $160   \pm 33.6$ & BC03 \cite{Moresco:2015cya} \\
    1.965  & $186.5 \pm 50.4$ & BC03 \cite{Moresco:2015cya} \\
    2.33   & $224  \pm 8  $   & BAO \cite{Bautista:2017zgn} \\
    \hline
    \hline
  \end{tabular}
  \begin{tabular}[t]{cc}
    \hline
    \hline
    $H(z)$ [km/s/Mpc] & Method \\
    \hline
    \\
    \\
    $81  \pm 5 $ & MaStro \cite{Moresco:2012jh} \\
    $81  \pm 6 $ & MaStro \cite{Moresco:2012jh} \\
    \\
    \\
    \\
    $88  \pm 16$ & MaStro \cite{Moresco:2012jh} \\
    \\
    \\
    \\
    \\
    \\
    \\
    \\
    $110 \pm 15$ & MaStro \cite{Moresco:2012jh} \\
    $98  \pm 10$ & MaStro \cite{Moresco:2012jh} \\
    $88  \pm 11$ & MaStro \cite{Moresco:2012jh} \\
    $124 \pm 17$ & MaStro \cite{Moresco:2012jh} \\
    \\
    $113 \pm 15$ & MaStro \cite{Moresco:2012jh} \\
    \\
    \\
    \\
    \hline
    \hline
  \end{tabular}
  \caption{Hubble parameter data obtained with BAO and galaxy
    differential age measurements. We use the model-independent
    determination $(1+z_*) r_\mathrm{s}=147.36\pm0.66$ Mpc to compute
    $H(z)$ given BAO constraints on $r_\mathrm{s} H(z)$
    \cite{Verde:2016wmz}.  Most differential age data are obtained
    assuming the BC03 stellar evolution model; we also consider data
    analysed with the MaStro stellar evolution model when available.}
  \label{tab:H}
\end{table}

For an observational determination $h(z)$, we rely on measurements of galaxy ages and the baryon acoustic oscillation (BAO) pattern imprinted on large scale structure.

With a measurement of galaxy ages $t(z)$, it is straightforward to
determine the expansion rate model-independently, as $1+z=a^{-1}$
implies $\frac{\rmd t}{\rmd z}=-\frac{1}{(1+z) H(z)}$
\cite{Jimenez:2001gg, Carson:2010sq, Zhang:2010ic}. There are now several
studies where $H(z)$ is determined from observations of passively
evolving galaxies \cite{Simon:2004tf, Stern:2009ep, Moresco:2012jh,
  Zhang:2012mp, Moresco:2015cya, Moresco:2016mzx}. We use the data of
\cite{Stern:2009ep, Moresco:2012jh, Zhang:2012mp, Moresco:2015cya,
  Moresco:2016mzx}, listed in \tab{tab:H}.\footnote{We do not use
  \cite{Simon:2004tf} due to concerns about the error determination
  \cite{Mortsell:2011yk}; see also the concerns of \cite{Crawford:2010rg}
  regarding single stellar population models, which are used in
  \cite{Zhang:2012mp}.} Some of the dating methods rely on global
spectral or photometric analysis \cite{Simon:2004tf, Stern:2009ep,
  Zhang:2012mp}, others use the 4000 \AA{} spectral break
\cite{Moresco:2010wh, Moresco:2012jh, Moresco:2015cya,
  Moresco:2016mzx}. The absolute age is not always well fit by the
models, but this is not a problem in itself, as $H(z)$ depends only on
the change of the age with redshift, and some systematic errors in the
absolute age cancel out in the differential age \cite{Moresco:2010wh,
  Moresco:2016mzx}\footnote{Ages inferred from the H$\b$ Lick index
  are also systematically too high, possibly due to unsubtracted
  emission lines, and there the uncertainty extends to redshift
  evolution, making it impossible to use that feature for
  determination of $H(z)$ \cite{Concas:2017bqi}.}.  At redshifts $z\gtrsim1$, the
differential age determination depends significantly on the adopted
stellar population synthesis model \cite{Moresco:2012jh}. Some authors have
chosen to add an extra 20\% error to high-$z$ datapoints to account
for this or drop some of the high-$z$ datapoints \cite{Verde:2014qea}. We
instead consider the results of different population synthesis models,
BC03 \cite{Bruzual:2003tq} and MaStro \cite{Maraston:2011sq}, and
compare. However, it should be kept in mind that comparison of
existing models (or simply increasing the error bars) does not
necessarily account for the systematic effects, which can only be
reliably settled by further study of the population synthesis
models. (A similar caveat applies to differences between SNIa
light-curve fitters.)

As the JLA dataset contains information only about the relative luminosity of SNe Ia, it can be used to obtain $d(z)$ without worrying about the absolute normalisation of the distance. In contrast, to compare the observed $H(z)$ data with the theoretical quantity $h(z)$, we also have to consider the normalisation factor $H_0$. While there are increasingly precise determinations of $H_0$ using local SNe \cite{Efstathiou:2013via, *Riess:2016jrr, *Cardona:2016ems, Zhang:2017aqn, Follin:2017ljs, Feeney:2017sgx}, they are in some tension with the value extrapolated from $H(z)$ determined from galaxy ages \cite{Busti:2014dua, Verde:2014qea, Busti:2014aoa}. We therefore keep $H_0$ as a free parameter in the fit.

The BAO pattern and its distortion by the Alcock--Paczy\'{n}ski effect can be used to measure $H(z)$ \cite{Alcock:1979mp, Bassett:2009mm, Montanari:2012me, Lepori:2016rzi}.  We use the determinations of $H(z)$ from the clustering of galaxies in SDSS-III BOSS \cite{Alam:2016hwk} and the clustering of quasars in the Ly$\a$ forest of SDSS DR12 \cite{Bautista:2017zgn}, listed in \tab{tab:H}.  The BAO observations constrain $r_\mathrm{s} H(z)$, where $r_\mathrm{s}$ is the sound horizon at decoupling. We use the model-independent determination $(1+z_*) r_\mathrm{s}=147.36\pm0.66$ Mpc (68\% C.I.), with $z_*=1090$ \cite{Verde:2016wmz}. The BAO data are not subject to similar astrophysical uncertainties as the SNIa and galaxy age data\footnote{However, see \eg \cite{DiDio:2016kyh} for an example of a possible systematic error due to inaccurate modelling that may be a concern for more precise future observations.}, but the analysis is more model-dependent. The study is usually carried out in terms of non-observable comoving coordinates, requiring the assumption of a fiducial cosmology. While correction parameters are introduced to account for possible deviations from the fiducial model \cite{Xu:2012fw}, it is not clear how well they describe models that are far from $\Lambda$CDM.  BAO analyses also rely on \LCDM mocks and the reconstruction technique applied to improve the statistics assumes $\Lambda$CDM, so the precise error bars should be treated with caution. (See \cite{Gaztanaga:2008xz, *MiraldaEscude:2009uz, *Labini:2009ke, *Kazin:2010nd, *Cabre:2010bc} for discussion of early BAO observations.)

Given the systematic errors of the cosmic clocks and model-dependence of BAO measurements, we compare results from different data combinations. The differences are non-negligible, but within the large errors the overall conclusions are similar.

\section{Fitting methods} \label{sec:method}

\subsection{Validation of numerical methods}

We fit $d(z)$ and $h(z)$ model-independently with polynomials and splines, and also consider the \LCDM model for comparison. Other model-independent data fitting procedures have been used in the literature, most notably Gaussian processes \cite{Seikel:2012cs, Verde:2014qea, Busti:2014dua, Busti:2014aoa, Cai:2015pia, Wei:2016xti, Yu:2016gmd}. However, the results of Gaussian processes depend on the choice of covariance function \cite{Seikel:2013fda, Busti:2014aoa} (see the discussion in appendix \ref{sec:red_corr}). Other possibilities include non-parametric smoothing \cite{Shafieloo:2009hi, LHuillier:2016mtc, Li:2016wjm, Gonzalez:2017tcm}, principal component analysis and genetic algorithms \cite{Sapone:2014nna} as well as radial basis functions \cite{Chiang:2017yrq}. Numerical derivatives and binning have also been considered \cite{Mortsell:2011yk, Sapone:2014nna}, but being able to take analytical derivatives significantly improves precision (although possibly at the cost of accuracy).

Our computations are implemented in the \textsc{Samp}\footnote{\textsc{Samp} (Samp's Adapted Monte Python) and other software used in this work are available under \emph{libre} licenses at \url{http://fmnt.info/projects/}.} code (based on \textsc{Monte Python} \cite{Audren:2012wb}). Curve fits are performed through Markov chain Monte Carlo (MCMC) methods to reconstruct Bayesian posteriors \cite{Trotta:2008qt}, based on the Metropolis--Hastings algorithm. As a convergence diagnostic for the MCMC chains (besides visual inspection of the chains traces) we require the Gelman--Rubin coefficient \cite{Gelman:1992zz} to be $R<1.1$. As a goodness-of-fit criterion we examine the reduced $\chi^2/dof$ (with $dof = N - n -1$, given $N$ data points and $n$ fitting parameters) and, in the spline case, a cross-validation analysis (see appendix \ref{sec:rough}). Unless stated otherwise, we consider flat uninformative priors except for the physical condition $\Omn>0$ in the $\Lambda$CDM case.  The numerical methods are validated through a two-fold process:

\begin{enumerate}

\item Unit tests are implemented for all main \textsc{Samp} functions
  related to polynomial and spline algorithms, prior to implementation
  into a specific likelihood code.

\item The likelihood code is tested by fitting mock data
  obtained replacing the observed data points by corresponding
  ones for a spatially flat $\Lambda$CDM model with $\Omn=0.3$ and $H_0=70$
  km/s/Mpc (with the same errors as in the real data). Such an
  analysis is repeated for each case.

\end{enumerate}

For the study of consistency relations based on SNIa light-curve data, we also implemented the computation of the maximised profiles (\ie the likelihood maximised in each MCMC histogram bin, also called a profile likelihood \cite{Planck:2013nga}) in \textsc{Samp}. Since MCMC methods may provide a poor description of the maximised profiles\footnote{A more easily computed quantity in MCMC analysis is the mean likelihood within each bin. However, it is not of relevant statistical interest aside from being a check of the robustness of results.}, we first compared a few results to those obtained with the robust Minuit minimisation algorithm \cite{James:1975dr}. More specifically, we used the Migrad algorithm included in the \textsc{iminuit}\footnote{\url{https://pypi.python.org/pypi/iminuit}} Python package, based on \textsc{SEAL Minuit2}.\footnote{\url{http://seal.web.cern.ch/seal/work-packages/mathlibs/minuit/}} Results for the likelihood best fits are recovered well within the $68\%$ confidence intervals, which are themselves consistent at the $\mathcal{O}(10\%)$ level. This uncertainty on the errors does not affect our conclusions, hence the maximised profile analysis is carried out in \textsc{Samp} based on the MCMC chains, at the same time as the marginalised posterior analysis.

\subsection{Supernova light-curve likelihood} \label{sec:curve}

Following the JLA analysis \cite{Betoule:2014frx}, the SNe Ia distance modulus is modelled as
\begin{equation}
  \mu(z) = m^*_B(z) - M_B + \alpha_x X_1(z) - \beta C(z) \;
\end{equation}
where $\alpha_x$ and $\beta$ are constants, and the $B$-band peak magnitude $m^*_B(z)$ and the Phillips colour $C(z)$ and stretch $X_1(z)$ corrections are obtained by fitting SN light-curves to photometric data, without having to specify the cosmology. To take into account dependence on host galaxy properties, the absolute magnitude $M_B$ is written in terms of the host stellar mass $M_{\mathrm{stellar}}$ as
\begin{equation}
  M_B = M^1_B +  \Delta_M \Theta(M_{\mathrm{stellar}}-10^{10}M_{\odot}) \;,
\end{equation}
where $M^1_B$ and $\Delta_M$ are constants and $\Theta(x)$ is the Heaviside step function.

The MCMC algorithm recovers the posterior distribution given the likelihood $\mathcal{L}_{\rm SN} \propto \exp\left( -\chi_{\rm SN}^2/2 \right)$, where
\begin{equation} \label{eq:chi2_jla}
  \chi_{\rm SN}^2 = \left( \hat{\boldsymbol\mu} - {\boldsymbol\mu}  \right)^T
  \textbf{C}^{-1} \left( \hat{\boldsymbol\mu} - {\boldsymbol\mu}  \right) \;,
\end{equation}
where $\hat{\boldsymbol\mu}$ represents the JLA data vector (each
component being identified by the corresponding redshift $z_i$) and
the model is written in terms of the luminosity distance $D_L$ as
\begin{equation}
  \mu_i = 5\log_{10}\left[ D_L({\boldsymbol \theta}, z_i)/(10 \ {\rm pc}) \right] = 5\log_{10}d_L({\boldsymbol \theta}, z_i) + M_{H_0} \;,
\end{equation}
where we have introduced $M_{H_0} \equiv -5 \log_{10}\left( 10 \ {\rm pc\ } H_0 \right)$. For discussion of the contributions to the covariance matrix $\textbf{C}$, see \cite{Betoule:2014frx}.  The fit constrains the parameters ${\boldsymbol \theta}$ that determine the functional form of the luminosity distance $d_L({\boldsymbol \theta}, z_i)$ (polynomial, spline or $\Lambda$CDM). As the JLA data contain no information about absolute luminosity, the parameter $H_0$ (which gives the normalisation of the luminosity distance $D_L=d_L/H_0$) used in the fit is fully degenerate with the absolute magnitude $M^1_B$. Without loss of generality, we fix $H_0=70$ km/s/Mpc (corresponding to $M_{H_0}\approx43.2$) and marginalise over $M^1_B$. (Note that this does not imply anything about the value of the physical Hubble parameter.) We are left with a total of four nuisance parameters $\alpha_x$, $\beta$, $\Delta_M$ and $M^1_B$, which are constant over the sample and marginalised over. In practice, we find that $\alpha_x$, $\beta$ and $\Delta_M$ depend only weakly on the cosmological model, and the values agree with \cite{Betoule:2014frx}. The values of $\a_x$ and $\b$ agree with the ones in \cite{Marriner:2011mf, Wei:2016xti, Li:2016wjm}, and $\a_x$ is slightly smaller than found for nearby SNe in \cite{Zhang:2017aqn}.

When fitting the polynomials and splines to the distance data, $d(z)$ has to satisfy certain physical conditions. We have the initial conditions
\begin{eqnarray} \label{eq:dic}
  d(z=0)=0 \ , \quad d'(z=0)=1 \;.
\end{eqnarray}
In addition, the Sachs equation \re{Sachs} shows that $\frac{\rmd^2d_A}{\rmd\l^2}<0$. This means that only those parameters that lead to a distance with at most one extremum (which is a maximum) are physical. (Note that this does not imply $\frac{\rmd^2d_A}{\rmd z^2}<0$.) The condition $\frac{\rmd^2d_A}{\rmd\l^2}<0$ is equivalent to the statement that $h(z)$ solved from the Sachs equation \re{hbr} is real, $h(z)^2>0$. These physical conditions are verified in \textsc{Samp} likelihood code on a fine redshift grid.

\subsection{Expansion rate likelihood}

When fitting to the expansion rate data, the posterior is recovered given the likelihood
$\mathcal{L}_h \propto \exp\left( -\chi_h^2/2 \right)$, with:
\begin{equation}
  \chi^2_h = \sum_{i=1}^{N} \frac{\left[{\hat
        h}(z_i)-h(z_i)\right]^2}{\sigma_{h_i}^2} \;,
\end{equation}
where ${\hat h}(z_i)$ denotes the expansion rate data at a given
redshift $z_i$, $\sigma_{h_i}$ are the respective errors and we have the boundary condition
\begin{equation} \label{eq:ic_h}
  h(z=0)=1 \;.
\end{equation}
The data provide ${\hat H}(z)=H_0{\hat h}(z)$, so we also vary $H_0$ as a free parameter. (This $H_0$ corresponds to the physical expansion rate today.)

\subsection{Fitting functions}

\subsubsection{Polynomial}

We use a fourth order polynomial to fit the luminosity distance $d_L(z)$:
\begin{equation} \label{eq:dL_polyfit}
  d_L(z, \boldsymbol{\theta}) = z+\theta_2z^2+\theta_3z^3 +\theta_4z^4 \;,
\end{equation}
where the coefficients $\theta_j$ are varied to fit the JLA distance modulus $\mu(z)$.  The homogeneous and linear coefficients ($\theta_0=0$ and $\theta_1=1$) are fixed by the initial conditions \re{eq:dic}. The polynomial order is set through mock data analysis, selecting the lowest order that provides fits accurate enough for our purposes. We tested with mocks also fitting a polynomial to $d_A(z)$ or $d(z)$ instead, but the luminosity distance proved to be the most accurate. (For discussion of different distances, see \cite{Cattoen:2007id, *Visser:2009zs}.)

The Hubble parameter $h$ is modelled with a second order polynomial:
\begin{equation} \label{eq:h_polyfit}
  h(z, \boldsymbol{\theta}) = 1 + \theta_1z + \theta_2z^2 \;.
\end{equation}
The homogeneous $\theta_0=1$ coefficient is set by the initial condition $h(z=0)=1$. Again, the order of the polynomial has been chosen to be the lowest that provides sufficiently accurate fits to mock data.

\subsubsection{Spline}

To fit spline functions we first set $n+1$ knots $z_k$, where $k=0,\ldots,n$, with $z_0=0$ and the rest logarithmically distributed over the data redshift range.  The corresponding luminosity distance values $d_L(z_k)$ are free parameters, except for $d_L(z_0)=0$.  We interpolate the points $\left\{z_k,y_k\right\}$, where $y_k\equiv d_L(z_k)$, with cubic splines so that the distance at a given redshift $z$ (within the interpolation range) is given by the spline algorithm:
\begin{equation}
  d_L (z) = {\rm
    spline}(z, \left\{z_k,y_k\right\}) \;.
\end{equation}
The parameters of the fit are $\theta_k=y_k$ for $k=1,\ldots,n$. We impose the boundary condition $d_L'(z_0)=1$ on the derivative at the first knot, as required by (\ref{eq:dic}), and marginalise over $d_L'(z_n)$ at the last knot. This last requirement leads to a better determination of the first derivative of the distance at large redshifts compared to the polynomial fit case. For the Hubble parameter, we take $h(z_0) = 1$ and set second derivatives at the boundaries using natural spline conditions $h''(z_0)=h''(z_n)=0$.\footnote{Marginalising over the derivatives at the Hubble parameter boundaries would only slightly increase errors without affecting our results. In contrast, arbitrarily fixing the derivative of the distance at the end point would introduce a non-negligible systematic bias, hence we marginalise over it.} The rest of the analysis is done in the same way as for the distance fit.

The number $n+1$ of spline knots must be large enough to allow a good fit, but sufficiently small to avoid overfitting.  Based on mock data analysis, we set $n+1=4$ for the luminosity distance and $n+1=3$ for the Hubble parameter. Increasing the number of knots does not improve the goodness-of-fit. Setting the knots on a linear (as opposed to logarithmic) scale in redshift would require a larger number of knots to reach a comparable fit. As a further check against overfitting, we also introduce a roughness parameter, whose value is set by a cross-validation analysis, described in more detail in appendix \ref{sec:rough}. The results indicate that the selected number of spline knots does not overfit the supernova nor the Hubble parameter data.

\subsubsection{$\Lambda$CDM}

In the $\Lambda$CDM model, the Hubble parameter at late times is
\begin{equation}
  h(z, \boldsymbol{\theta}) = \sqrt{\Omn (1+z)^3 +
    \Omega_{K0}(1+z)^2 + \Omega_{\Lambda0}} \;,
\end{equation}
with $\Omega_{\Lambda0}=1-\Omn-\Omega_{K0}$, and the
set of varying parameters $\boldsymbol{\theta}$ is $\Omn$ and $\Omega_{K0}$.

In this case, the distance can be solved from the Sachs equation \re{Sachs} in closed form,
\begin{equation}
  d(z, \boldsymbol{\theta}) = \frac{1}{\sqrt{\Omega_{K0}}}
  \sinh\left( \sqrt{\Omega_{K0}} \int_0^z \frac{d\tilde{z}}{h(\tilde z,\Omn,\Omega_{K0})} \right)
  \;.
\end{equation}

\subsubsection{Derived functions} \label{sec:der}

The polynomial, spline and $\Lambda$CDM fits provide the uncertainties on the respective parameters used to model the luminosity distance and the Hubble parameter. However, we are not interested in the fit parameters themselves, but on the error contours of functions of those parameters. Specifically, we want to find the distance $d(z)$, Hubble parameter $h(z)$,\footnote{Spline fits already directly provide distances and Hubble parameters at the respective redshift knots, but we are interested in recovering such functions (and others) at any arbitrary redshift within the data range.} consistency condition functions $k_H(z)$, $k_S(z_\ml, z_\ms)$, $k_P(z)$ and the equations of state $\wHtot(z)$ and $\wDtot(z)$. To estimate error contours, we compute a given function $f_i = f(z_i)$ (at a fixed redshift value $z_i$) for each point of the MCMC chains. Error contours are then easily computed at each redshift $z_i$ (in practice we only consider a few redshift values). In the interpretation of the results, it should be kept in mind that there are non-trivial correlations between different redshifts, as discussed in appendix \ref{sec:red_corr}.

Unless otherwise noted, we quote error bars as 68\% C.I. and limits as
95\% C.I., the expression ``C.I.'' referring to confidence intervals
for maximised statistics and minimum credible intervals for
marginalised statistics (see appendix \ref{sec:vol_effect} for details); the meaning should be clear from the context.

\begin{figure}[ht]
  \centering
  \includegraphics[width=0.7\textwidth]{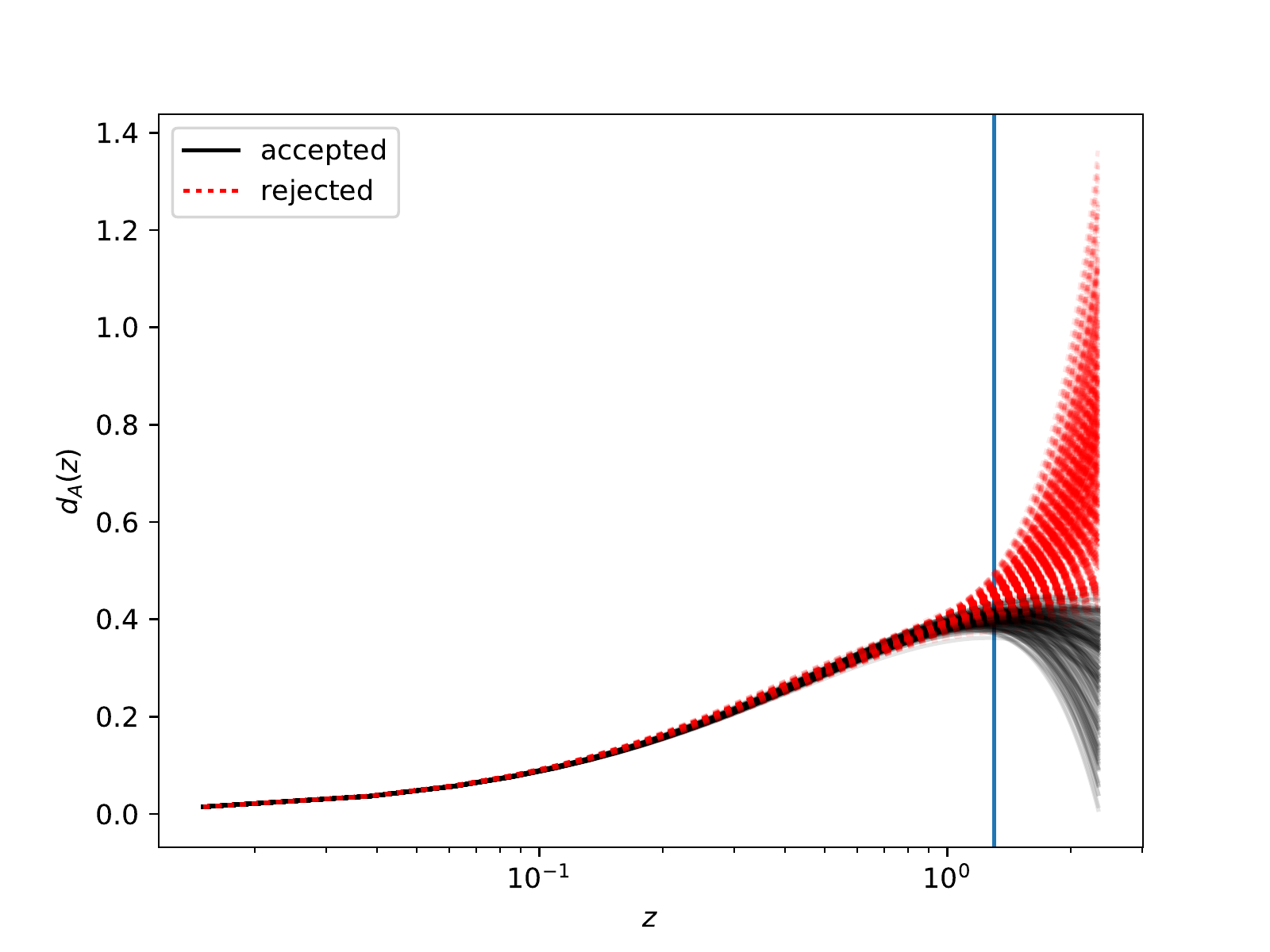}
  \caption{Angular diameter distances randomly extracted from a MCMC
    chain where a polynomial luminosity distance is fitted to the JLA
    SNIa data without implementing the condition that $h(z)$ is
    non-singular. The vertical line marks the last JLA datapoint.}
  \label{fig:rejected_dA}
\end{figure}

\subsection{Distance selection bias} \label{sec:bias}

When we determine the backreaction expansion rate from \re{hbr}, we are faced with the issue that the denominator of \re{hbr} vanishes at $z_\mathrm{m}$ where $d'_A=0$, as noted earlier. Therefore, for $d_A$ curves with a maximum, the value of $\Omn$ has to be chosen such that the numerator vanishes at the same redshift. This leads to two problems.

First, $z_\mathrm{m}$ is typically somewhat larger than the maximum redshift in the data, so the $\Omn$ value relies on extrapolating the fit. Spline fits cannot be extrapolated, so we only consider polynomials here. The extrapolation is not large, as $z_\mathrm{m}$ is typically not far outside of the data range (in the spatially flat \LCDM model with $\Omn=0.3$, the value is $z_\mathrm{m}=1.6$), and mock data analysis suggests that the recovered $\Omn$ values are consistent with the fiducial cosmology. Fitting Union2.1 data, which go to slightly larger redshifts ($z<1.4$) than JLA ($z<1.3$) or adding a distance data point at $z=2.34$ from Ly$\alpha$ BAO analysis \cite{Delubac:2014aqe} would not substantially affect the conclusion, while increasing model dependence.

Second, and more important, not all curves that provide a good fit to the data have a maximum, as shown in figure~\ref{fig:rejected_dA}.
We reject such curves (some of them would lead to $h^2<0$, a symptom of not satisfying the condition $\frac{\rmd^2 d_A}{\rmd\l^2}<0$). Redshifts beyond the JLA range ($z<1.3$) are not considered in the $\chi^2$, but we extend the search for the maximum up to $z=2.34$. This is a good compromise between including well-fitting curves with $d'_A=0$ but not considering too large redshifts, where the extrapolation of the polynomial becomes unreliable. The cutoff redshift is somewhat arbitrary, but the results are not very sensitive to the its precise value; for example, choosing $z=2$ instead does not change the results.  Mock data analysis shows that the selection effect of demanding $d_A$ to have a maximum in this range causes significant bias in the marginalised statistics, but not in the maximised statistics. We discuss this in more detail in appendix \ref{sec:vol_effect}. If we did not demand $d_A$ to have a maximum, some of the curves would still have to be discarded because they would lead to $h^2<0$.

In summary, when determining the backreaction case $h(z)$ from the distance data, we require $d_A$ to have exactly one maximum (in order to set $\Omn$ such that $h^2$ is positive and non-singular), and focus on maximised rather than marginalised profiles.

\begin{figure}[ht!]
  \centering

  \begin{subfigure}[b]{0.48\textwidth}
    \includegraphics[width=\textwidth]{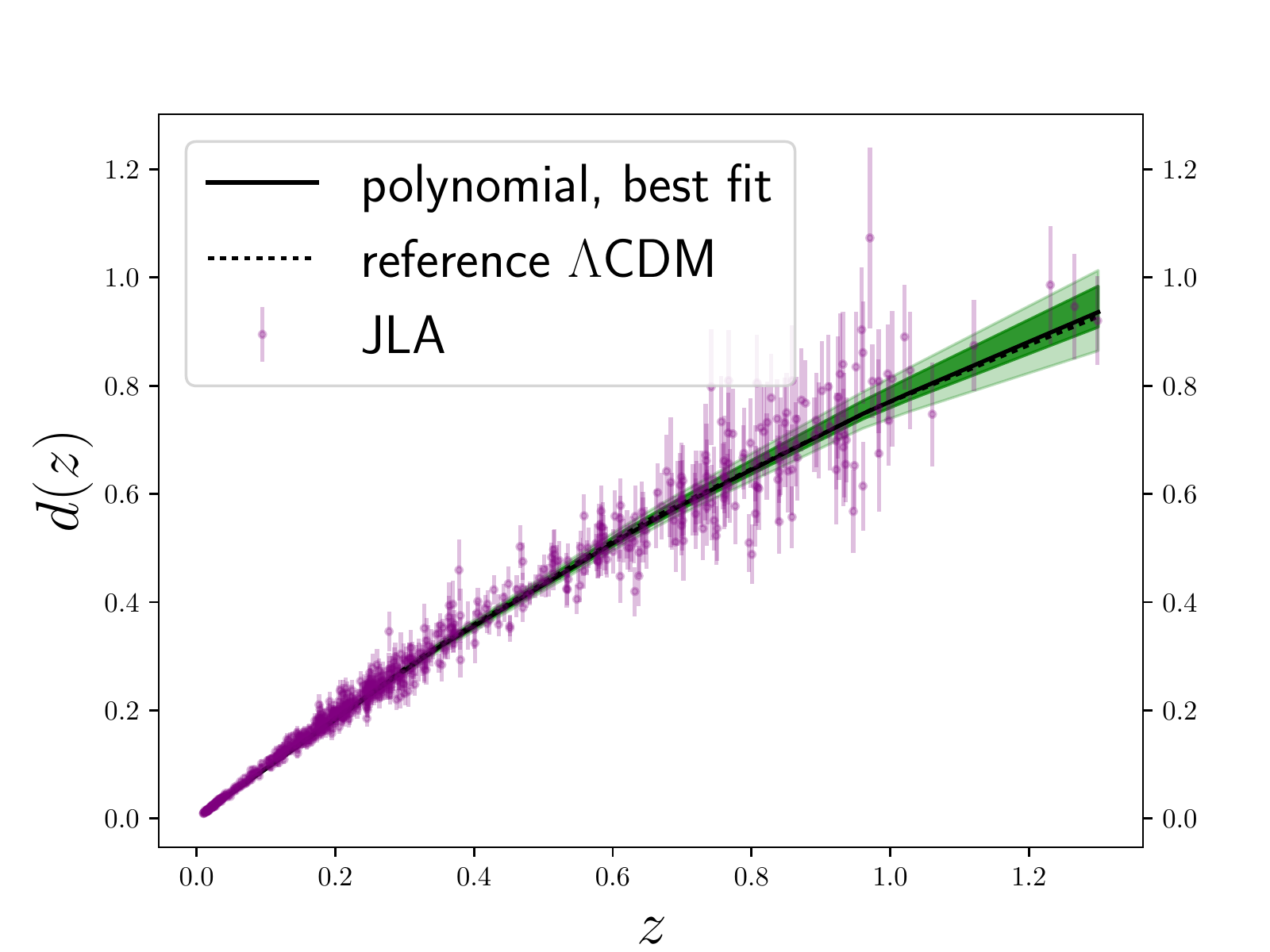}
    \caption{}
    \label{fig:d_BestFit_jla_poly_sachs}
  \end{subfigure}
  ~
  \begin{subfigure}[b]{0.48\textwidth}
    \includegraphics[width=\textwidth]{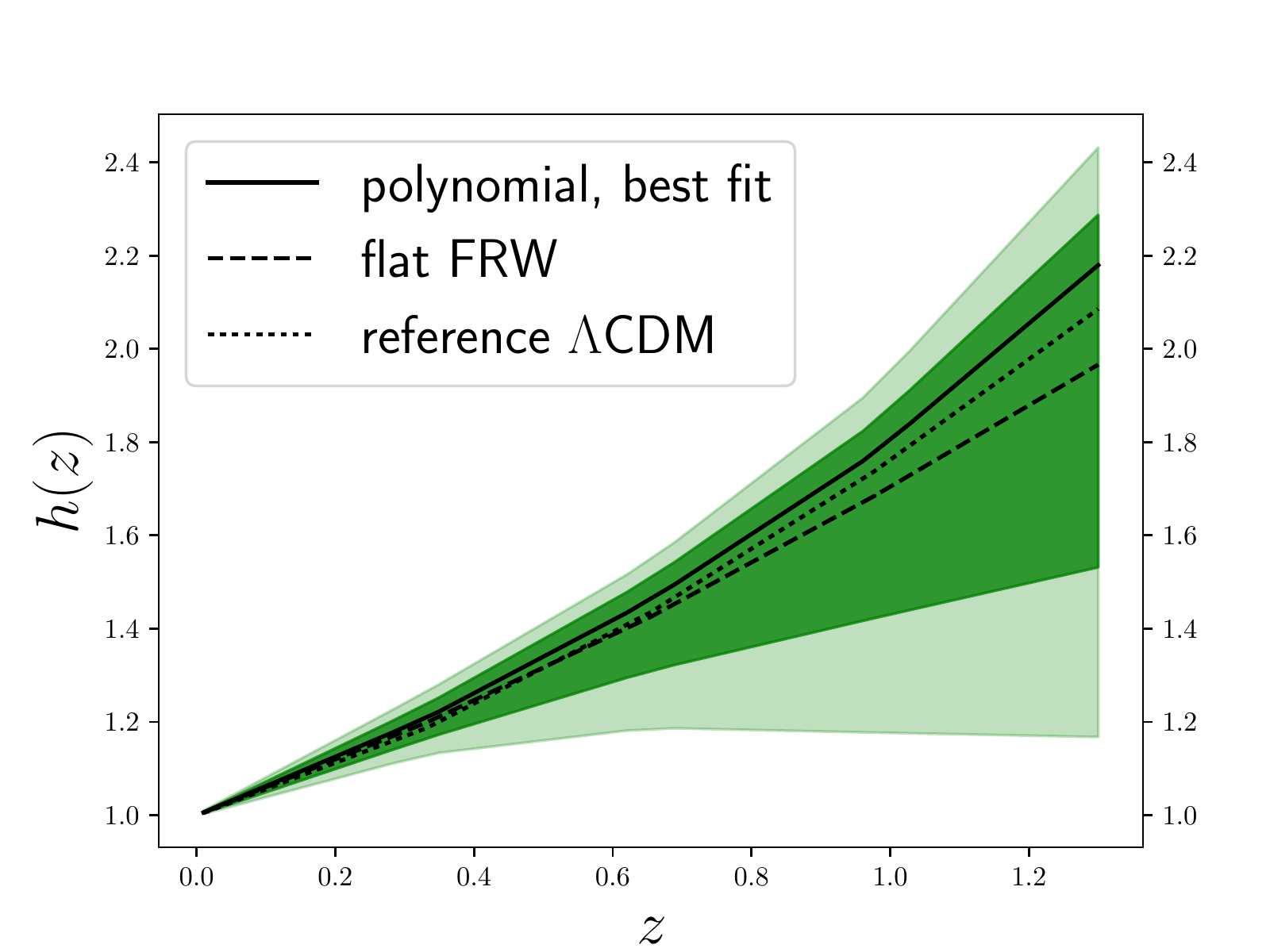}
    \caption{}
    \label{fig:hz_BestFit_poly_sachs}
  \end{subfigure}
  ~
  \\

  \begin{subfigure}[b]{0.48\textwidth}
    \includegraphics[width=\textwidth]{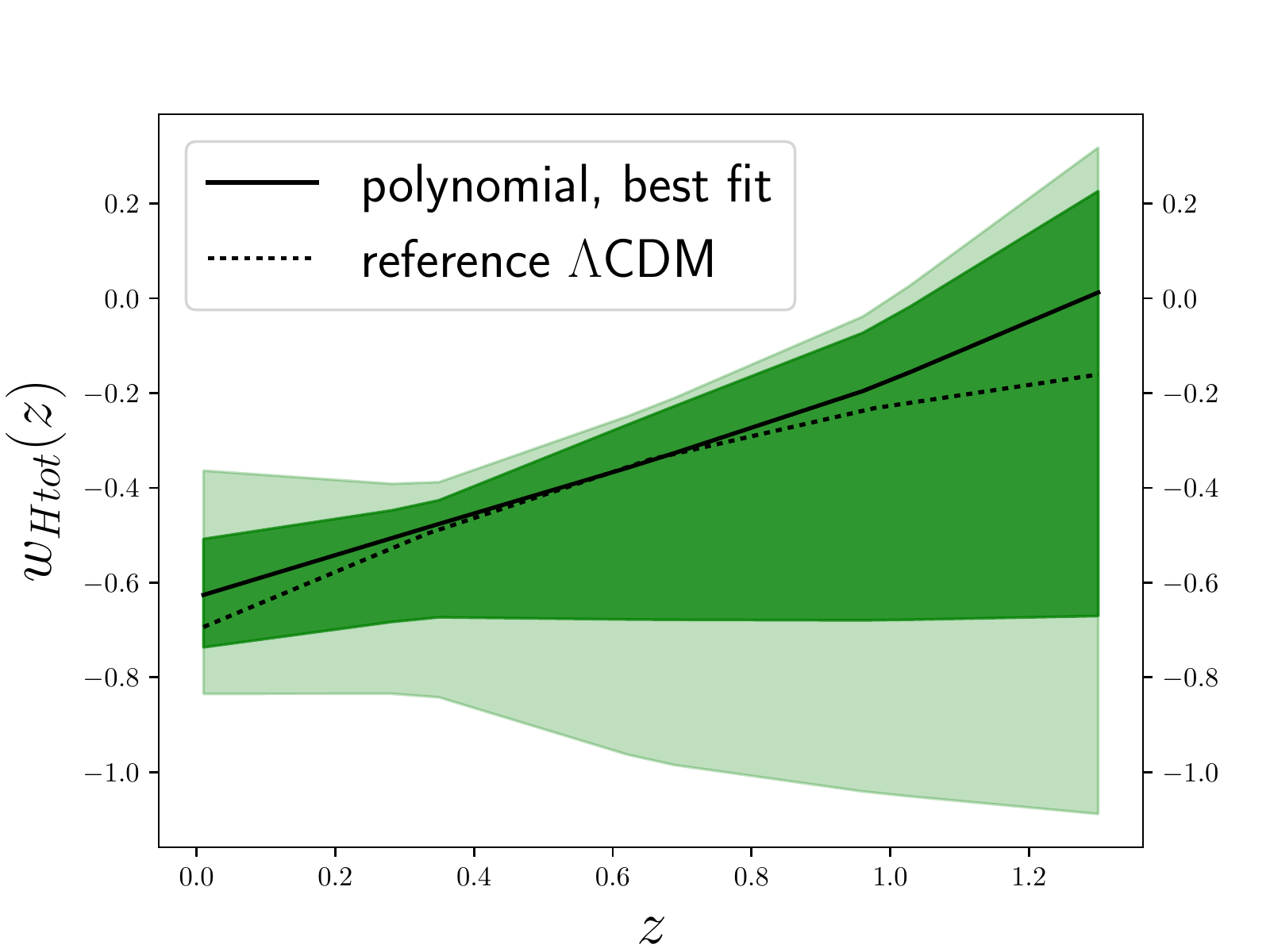}
    \caption{}
    \label{fig:wHtot_BestFit_jla_poly_sachs}
  \end{subfigure}
  ~
  \begin{subfigure}[b]{0.48\textwidth}
    \includegraphics[width=\textwidth]{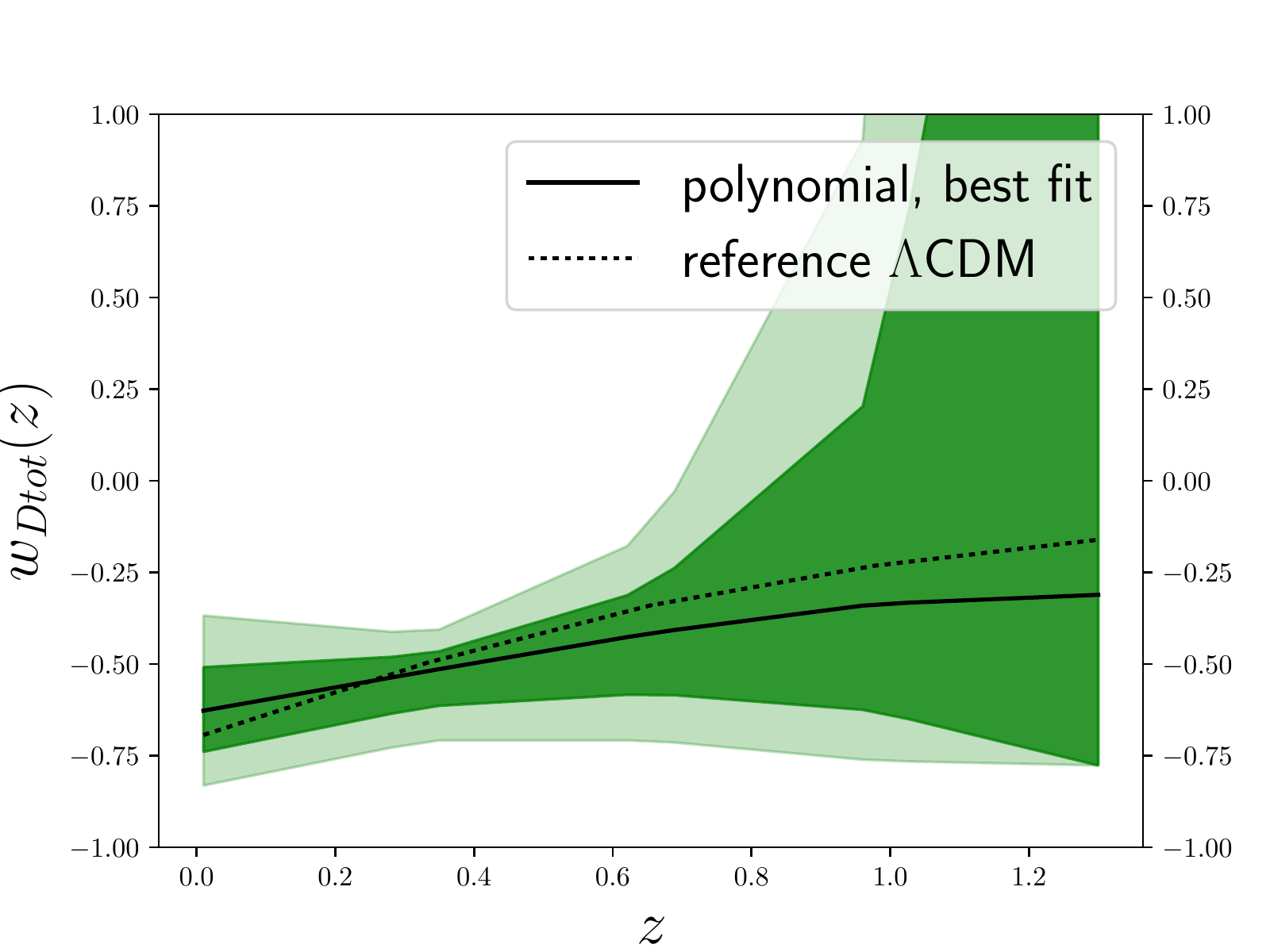}
    \caption{}
    \label{fig:wDtot_BestFit_jla_poly_sachs}
  \end{subfigure}
  ~
  \\

  \begin{subfigure}[b]{0.48\textwidth}
    \includegraphics[width=\textwidth]{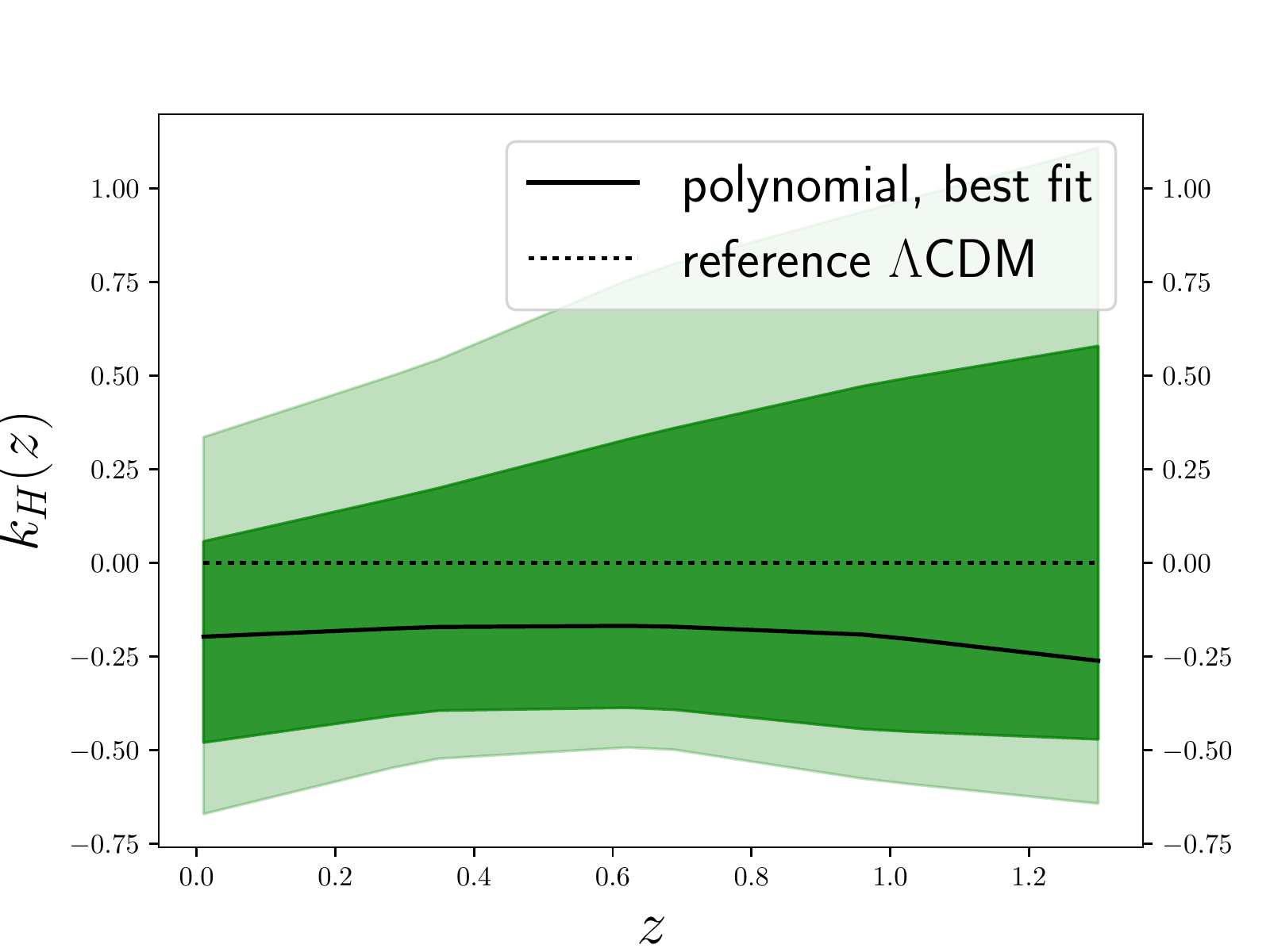}
    \caption{}
    \label{fig:kH_BestFit_jla_poly_sachs}
  \end{subfigure}
  ~

  \caption{Best fits and maximised 68\% and 95\% C.I. for the distance $d(z)$, the expansion rate $h(z)$, the equations of state $w_{H\textrm{tot}}(z)$ and $w_{D\textrm{tot}}(z)$, and the consistency condition function $k_H(z)$. In addition to the backreaction case, we show the \LCDM curve for comparison. For the expansion rate, we also show the $h(z)=1/d'(z)$ for the polynomial best fit distance, corresponding to the spatially flat FRW case.}
  \label{fig:jla_sachs}
\end{figure}

\section{Results} \label{sec:res}

\subsection{Expansion rate and consistency conditions from backreaction} \label{sec:frw_jla_sachs}

\subsubsection{Distance and expansion rate}

We first determine the luminosity distance from the JLA data, then find the corresponding backreaction $h(z)$ with \re{hbr} and calculate the allowed range for $k_H(z)$, $k_S(z_\ml,z_\ms)$ and $k_P(z)$.

Figure~\ref{fig:jla_sachs} shows the results obtained by fitting the polynomial luminosity distance to the JLA SNIa data.  The data points shown are obtained by fixing the nuisance parameters to their best fit values.  The distance is relatively well constrained: the polynomial best fit, shown in \fig{fig:d_BestFit_jla_poly_sachs}, is close to the \LCDM best fit, and the 95\% contours are quite tight, with only 10\% errors even at the largest redshift. The result for the matter density determined from the maximum of $d_A$, $\Omn=0.24_{-0.11}^{+0.12}$, is close to the  result for the \LCDM model fitted to the same JLA data, $\Omn=0.21_{-0.12}^{+0.10}$.  In contrast, the Hubble parameter determined from \re{hbr}, shown in \fig{fig:hz_BestFit_poly_sachs}, shows large errors at high redshifts, in part due to the spread in $\Omn$ values determined with the extrapolation.  In addition to the backreaction and \LCDM curves, we show the curve corresponding to the polynomial best fit distance in a spatially flat FRW model, where $h=1/d'$. It shows a clear difference from the backreaction case, of order 10\% at high redshift, suggesting significant violation of the FRW consistency condition $k_H$ (though it has to be checked whether this could be fitted by FRW spatial curvature). However, the difference is completely swamped by the large errors. The errors could be reduced by adding an angular diameter distance determination at $z=2.34$ from Ly$\alpha$ BAO analysis \cite{Delubac:2014aqe}, but this would increase model dependence without changing the main conclusions. The errors are smaller at low redshifts due to a better determination of the distance, combined with the initial conditions at $z=0$.

The uncertainty in the total equations of state $w_{H\textrm{tot}}$ and $w_{D\textrm{tot}}$, shown in figure~\ref{fig:jla_sachs}, is even larger, and they are consistent with each other. This is not surprising, as the equations of state depend on the second derivative of the distance, and even small errors in the distance can lead to large errors in the equation of state \cite{Fleury:2016fda}, as our mock studies confirm.  At large redshift, the errors are much larger for $w_{D\textrm{tot}}$ than for $w_{H\textrm{tot}}$. This is mainly due to the fact that the $\Omn$ values are determined to provide a finite $h(z)$ from \re{hbr}, which then enters in $w_{H\textrm{tot}}$. Instead, $w_{D\textrm{tot}}$ depends on $h=1/d'$, which is more affected by the spread in distance values introduced by the extrapolation.
While increasing the polynomial order leads to more precision in the derivatives, it does not necessarily increase accuracy due to overfitting and the fact that in general higher-order polynomials do not provide a more reliable extrapolation (used to determine $\Omn$).

\subsubsection{Backreaction prediction for $k_H(z)$}

We input the $d(z)$ determined from the JLA data and the corresponding $h(z)$ calculated from \re{hbr} into the expression \re{kH} for $k_H(z)$: the result is shown in figure~\ref{fig:kH_BestFit_jla_poly_sachs}. In contrast to the effective equations of state, $k_H$ depends only on the first derivative of $d$, though it is still affected by the spread in $\Omn$. The best fit shows a clear deviation from the spatially flat FRW case, with $k_H(z)\approx-0.25$ for the whole data range. This is the same order of magnitude as obtained in \cite{Boehm:2013qqa}, where a backreaction toy model was fitted to the data, but the redshift-dependence is completely different. The fit result is almost constant, reflecting the fact that the distance data does not require the kind of features in $h(z)$ that are expected if backreaction explains the accelerated expansion (such as early extra deceleration and possible late transition from acceleration back to deceleration). Because $k_H(z)$ is almost constant, it would be difficult to distinguish backreaction from FRW spatial curvature using observations at redshifts $z\lesssim1$. However, the behaviour at high redshifts would be completely different, as the FRW constant $k$ has a large impact on the distance to the last scattering surface (which is precisely constrained model-independently \cite{Vonlanthen:2010cd, *Audren:2013nwa, Audren:2012wb}) unlike a backreaction $k_H(z)$, which would be expected to rapidly decrease for $z\gg1$ as the effect of non-linear structures becomes negligible \cite{Boehm:2013qqa}. In any case, there is no detection of any deviation from the FRW case and within the error bars it is not possible to make out any trend of redshift evolution. The large errors still allow for features in $k_H(z)$, with the 95\% confidence contours covering the range $|k_H(z)|\sim1$, though because of large correlations between different redshifts (see appendix \ref{sec:red_corr} for details) it is not possible to put constraints on them simply by considering \fig{fig:kH_BestFit_jla_poly_sachs}. The tightest constraint is $-0.67<k_H(z)<0.34$ at $z=0.01$.

This result shows how much deviation from the FRW consistency condition is expected given the observed light-curves, if backreaction is responsible for the accelerated expansion and the relation between the average expansion rate and distance is given by the relation \re{hbr}. Before comparing this to what the combination of distance and expansion rate observations gives for $k_H(z)$, let us also find the backreaction range for $k_S(z_\ml, z_\ms)$ and $k_P(z)$.

\subsubsection{Backreaction prediction for $k_S(z_\ml, z_\ms)$}

\begin{figure}[t!]
  \centering

  \begin{subfigure}[b]{0.48\textwidth}
    \includegraphics[width=\textwidth]{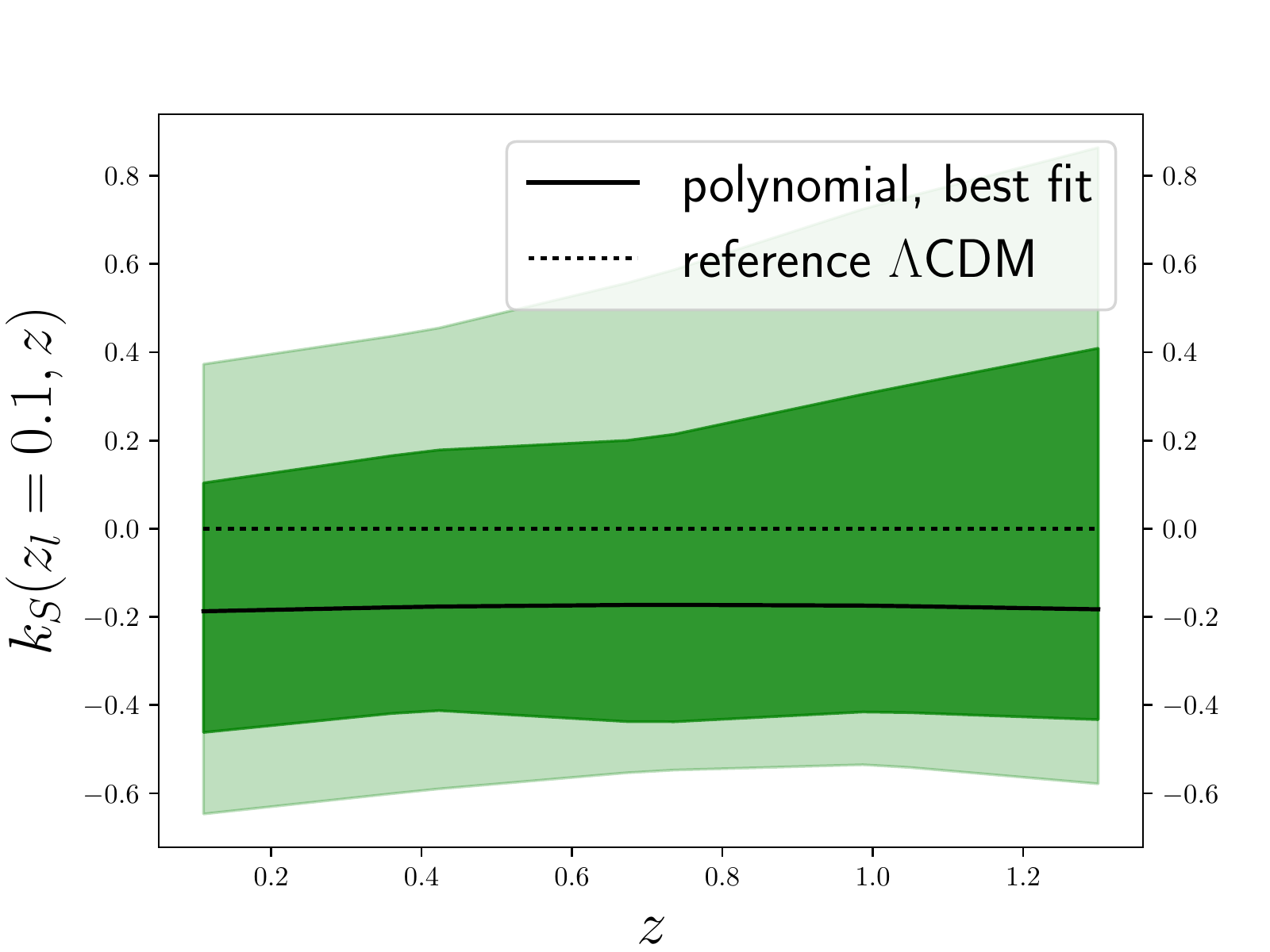}
    \caption{}
    \label{fig:kS_l0.1_BestFit_jla_poly_sachs}
  \end{subfigure}
  ~
  \begin{subfigure}[b]{0.48\textwidth}
    \includegraphics[width=\textwidth]{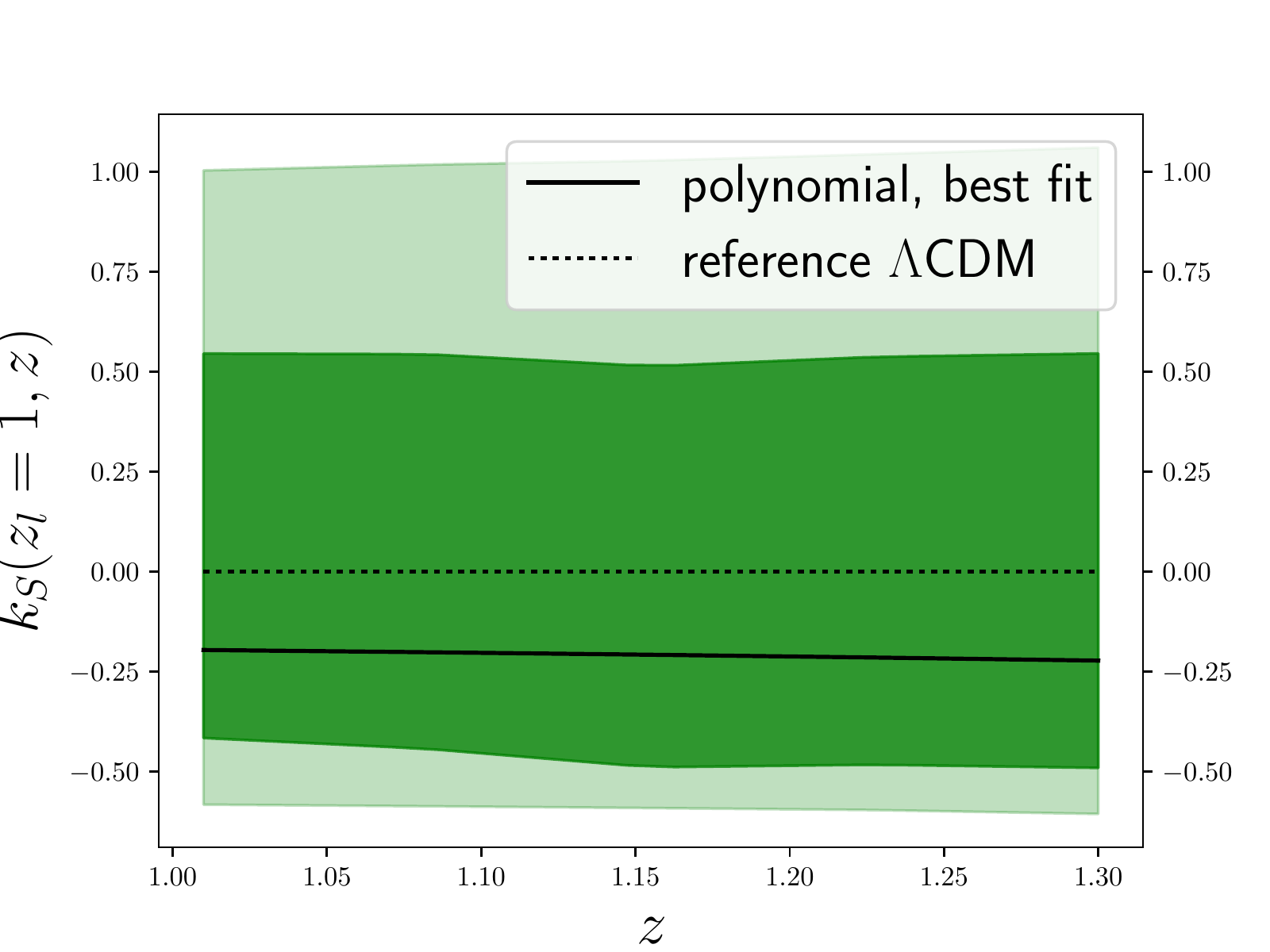}
    \caption{}
    \label{fig:kS_l1.0_BestFit_jla_poly_sachs}
  \end{subfigure}
  ~
  \\

  \caption{Best fit and the maximised 68\% and 95\% C.I. intervals for the backreaction prediction for
    $k_S(z_\ml, z_\ms)$ for $z_l=0.1$ and 1.0.}
  \label{fig:jla_sachs_kS}
\end{figure}

The backreaction sum rule consistency condition function $k_S(z_\ml, z_\ms)$ is determined from the polynomial fit to the observed $d(z)$ by first finding $h(z)$ from \re{hbr}, determining $d(z_\ml,z_\ms)$ from \re{dls} and inputting them together into \re{kS}. The result is shown in figure~\ref{fig:jla_sachs_kS}. In the two plots we show the cases where $z_\ml$ is 0.1 or 1.0. The best fit goes slightly down with $z_\ml$, and the errors increase with larger $z_\ml$, because the distance is more poorly determined. The dependence on $z_\ms$ is small: in both cases the best fit prediction is almost a straight line, the errors just increase slightly.

The allowed range is similar to the first observational constraints \cite{Rasanen:2014mca}, which can be much improved. (Note that in \cite{Rasanen:2014mca} it was assumed that $k_H$ is constant. If redshift dependence were allowed, the constraints would be wider, as we discuss in \sec{sec:dhdata} for $k_H$.)

\subsubsection{Backreaction prediction for $k_P(z)$}

\begin{figure}[t!]
  \centering
    \includegraphics[width=0.48\textwidth]{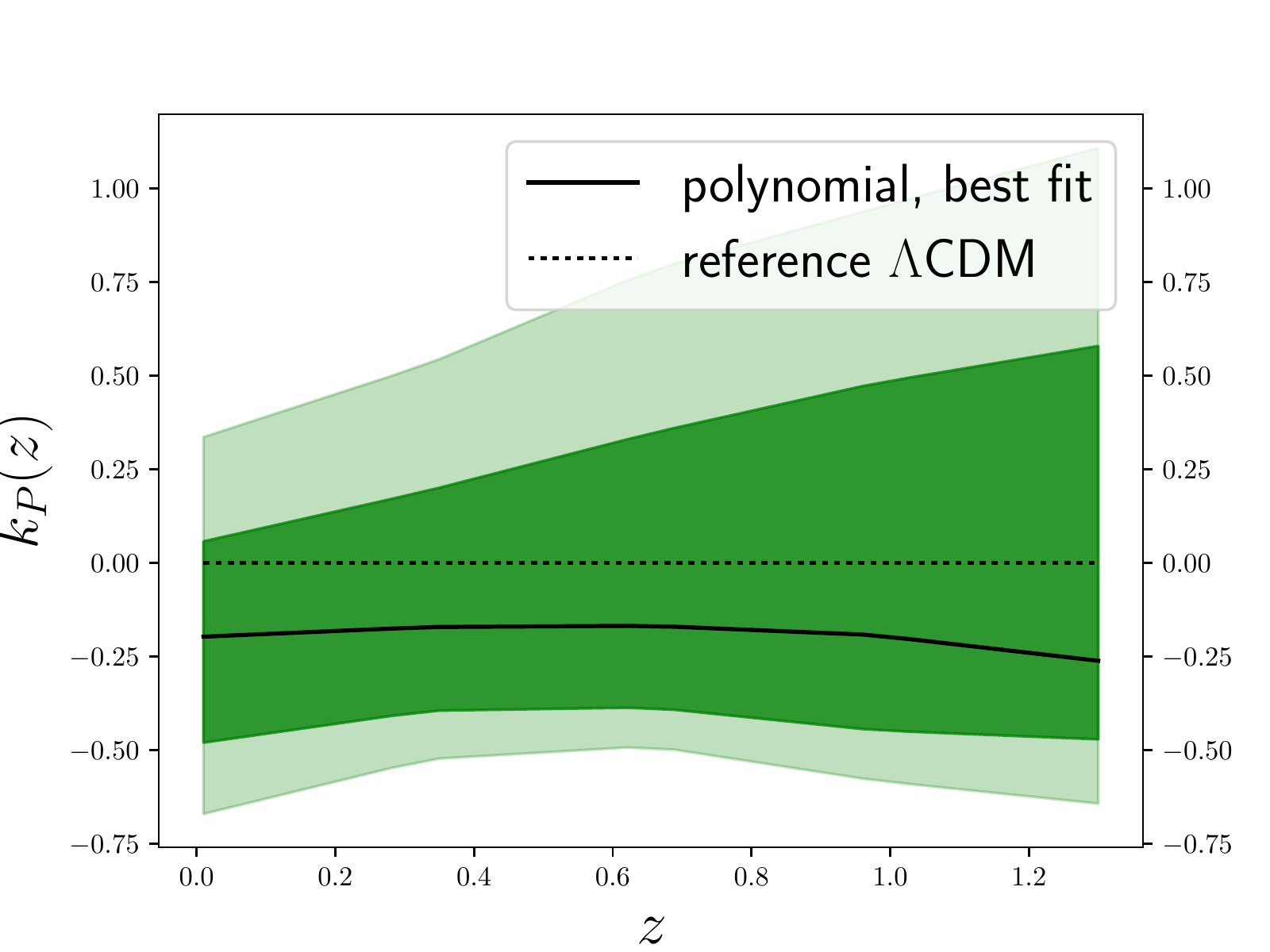}
     \caption{Best fit and the maximised 68\% and 95\% C.I. for the backreaction prediction for
    $k_P(z)$.}
  \label{fig:kP}
\end{figure}

As $k_P(z)=k_S(z,z)$, and the best fit $k_S(z_\ml,z_\ms)$ is close to constant in both redshift variables, this implies that the best fit $k_P$ is also close to constant, as we indeed see in \fig{fig:kP}. For the best fit, $k_P\approx k_S\approx-0.25\ldots-0.2$, and the error contours are similar as for $k_S$. If one of the consistency condition functions were strictly constant, they would all be constant and equal to each other, so it is not surprising that as they are (for the best fits) almost constant, they are also close to each other. In fact, the contours for $k_H$ and $k_P$ are essentially indistinguishable. However, the 95\% C.I. contours cover a wide range of possible values and redshift-dependencies.

There are no observations of the parallax distance on cosmological distances at the moment, but upcoming data from the Gaia satellite may yield a constraint of the order $|k_P|\lesssim1$, provided that there will be measurements of $d(z)$ at the corresponding redshifts \cite{Rasanen:2013swa}.

\subsection{Distance, expansion rate and consistency conditions from the data} \label{sec:dhdata}

\subsubsection{Fits to the distance data}
\label{sec:dhdata_dist}

\begin{figure}[t]
  \centering

  \begin{subfigure}[b]{0.48\textwidth}
    \includegraphics[width=\textwidth]{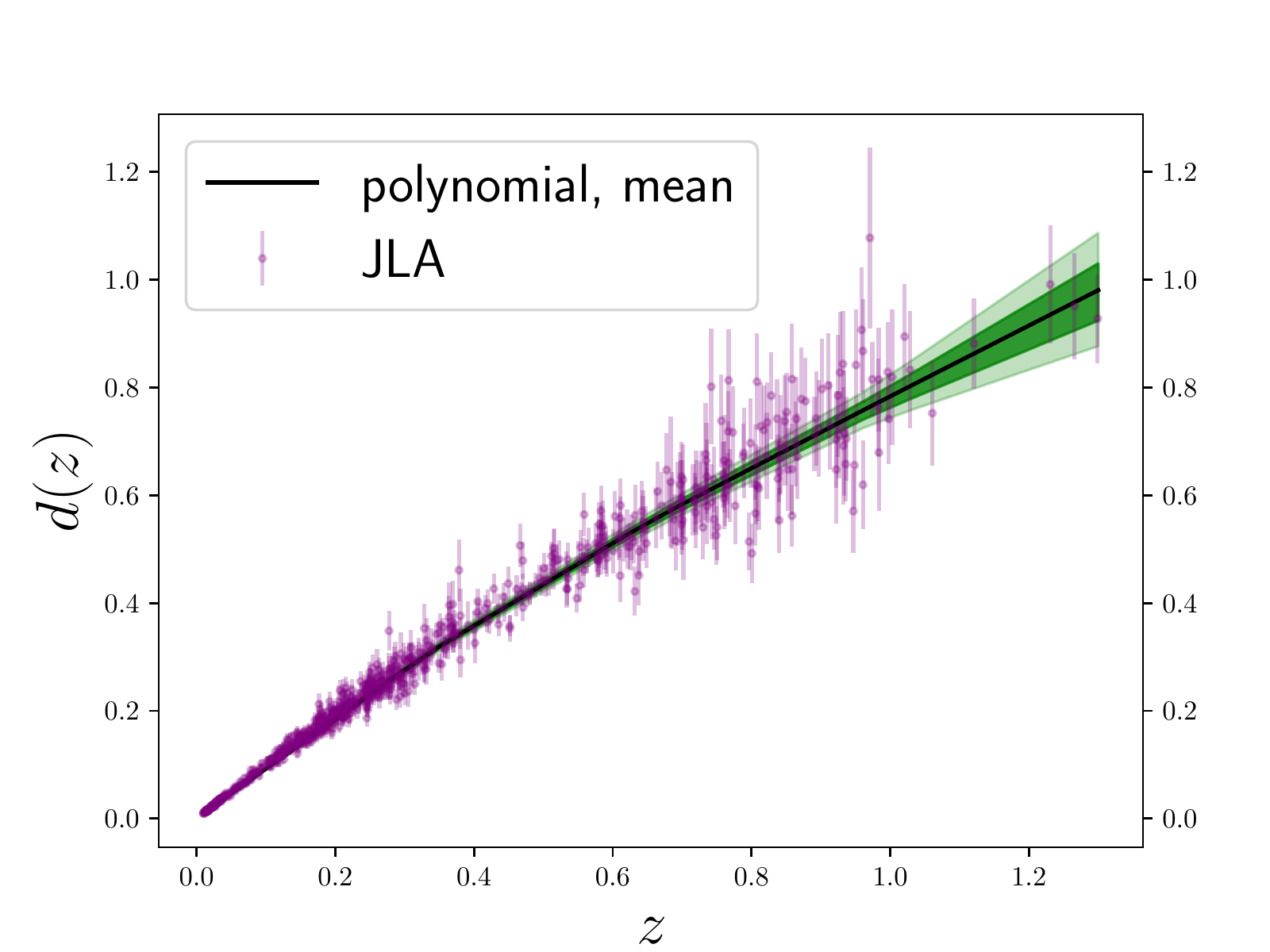}
    \caption{$\chi^2_\textrm{min}/dof=682/732$.}
    \label{fig:d_mean_jla_poly}
  \end{subfigure}
  ~
  \begin{subfigure}[b]{0.48\textwidth}
    \includegraphics[width=\textwidth]{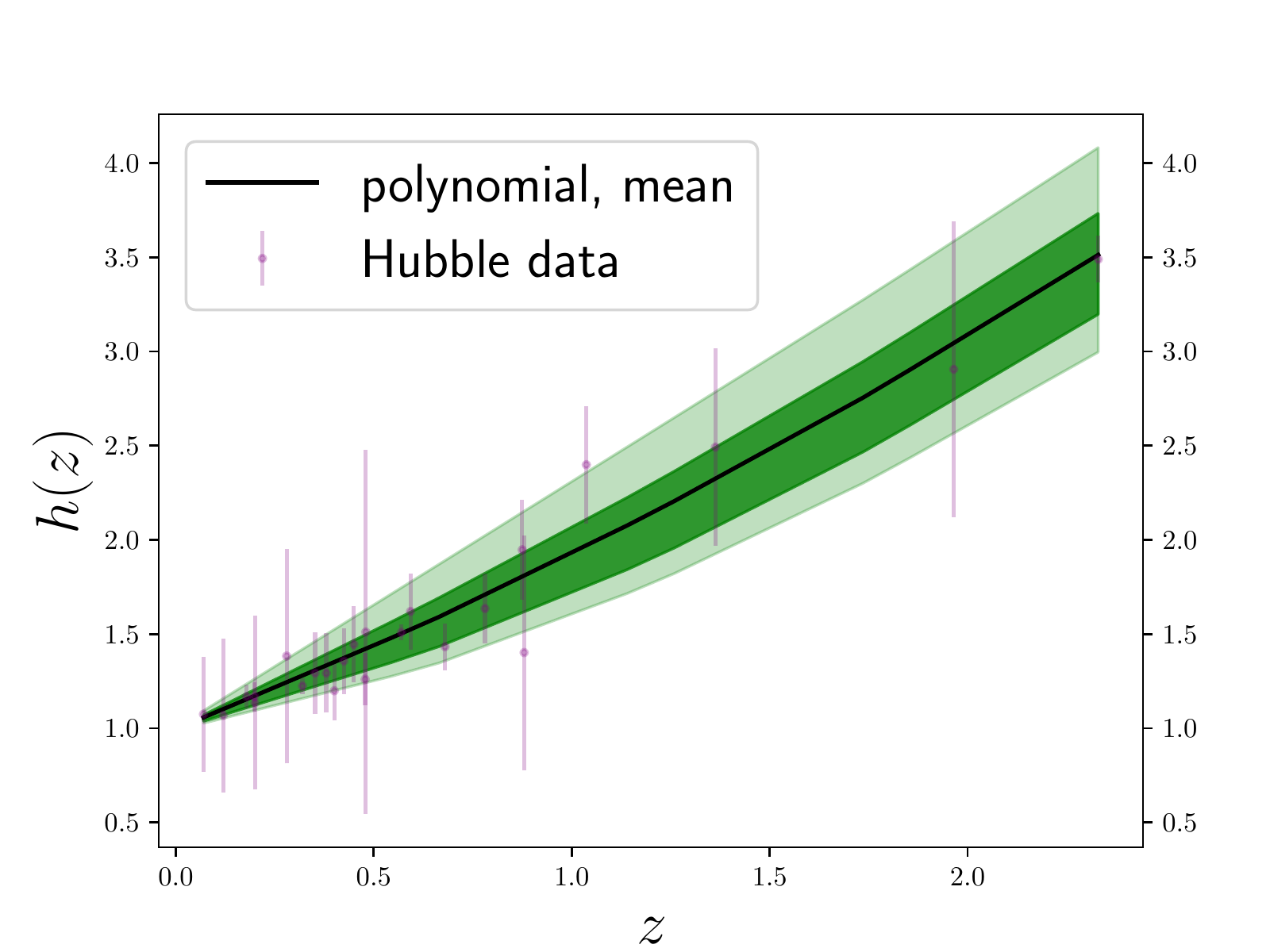}
    \caption{$\chi^2_\textrm{min}/dof=7/20$.}
    \label{fig:hz_mean_hubble_poly_BC03_BAO}
  \end{subfigure}
  ~
  \\

 \begin{subfigure}[b]{0.48\textwidth}
    \includegraphics[width=\textwidth]{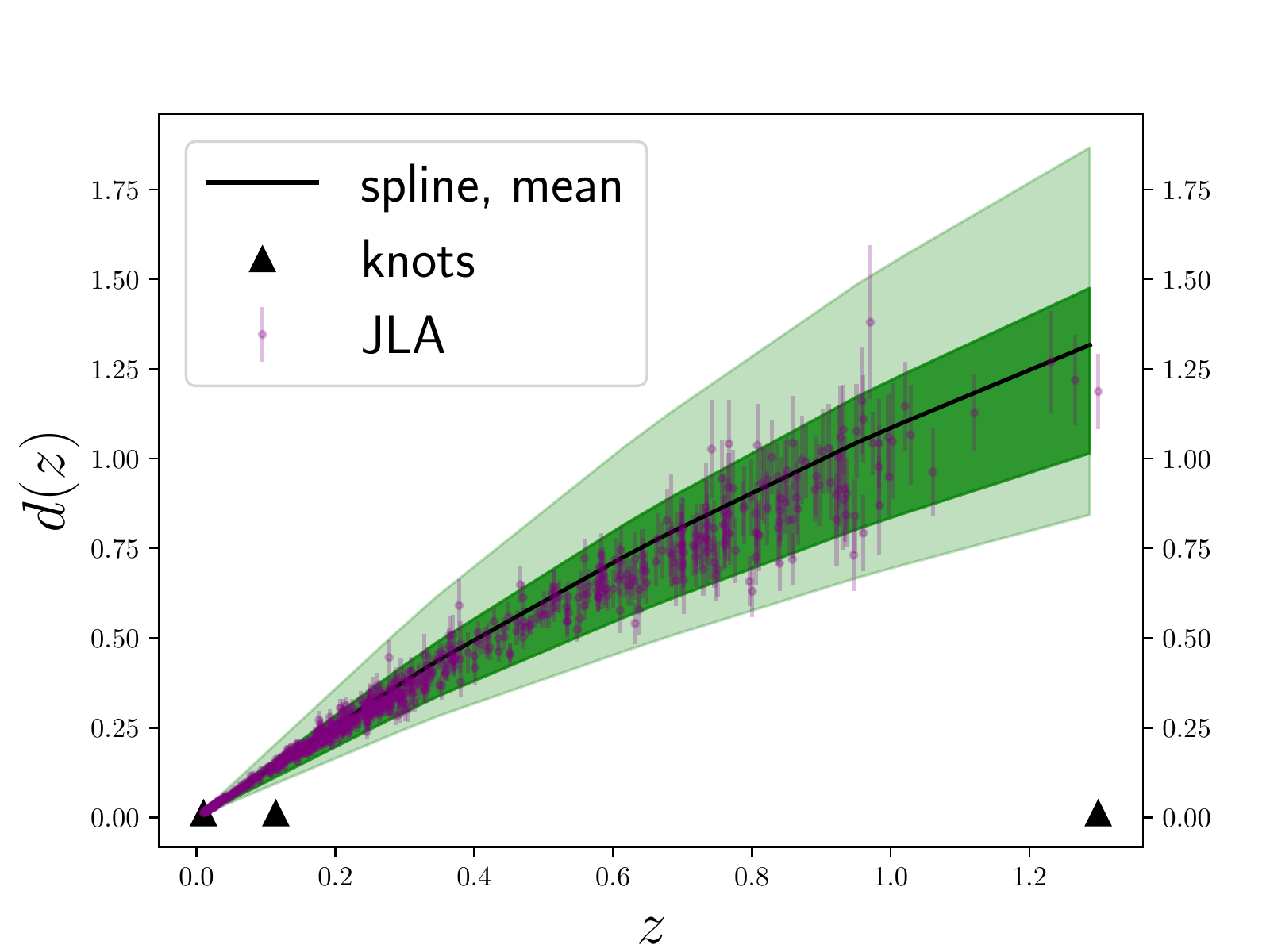}
    \caption{$\chi^2_\textrm{min}/dof=688/731$.}
    \label{fig:d_mean_jla_spline}
  \end{subfigure}
  ~
  \begin{subfigure}[b]{0.48\textwidth}
    \includegraphics[width=\textwidth]{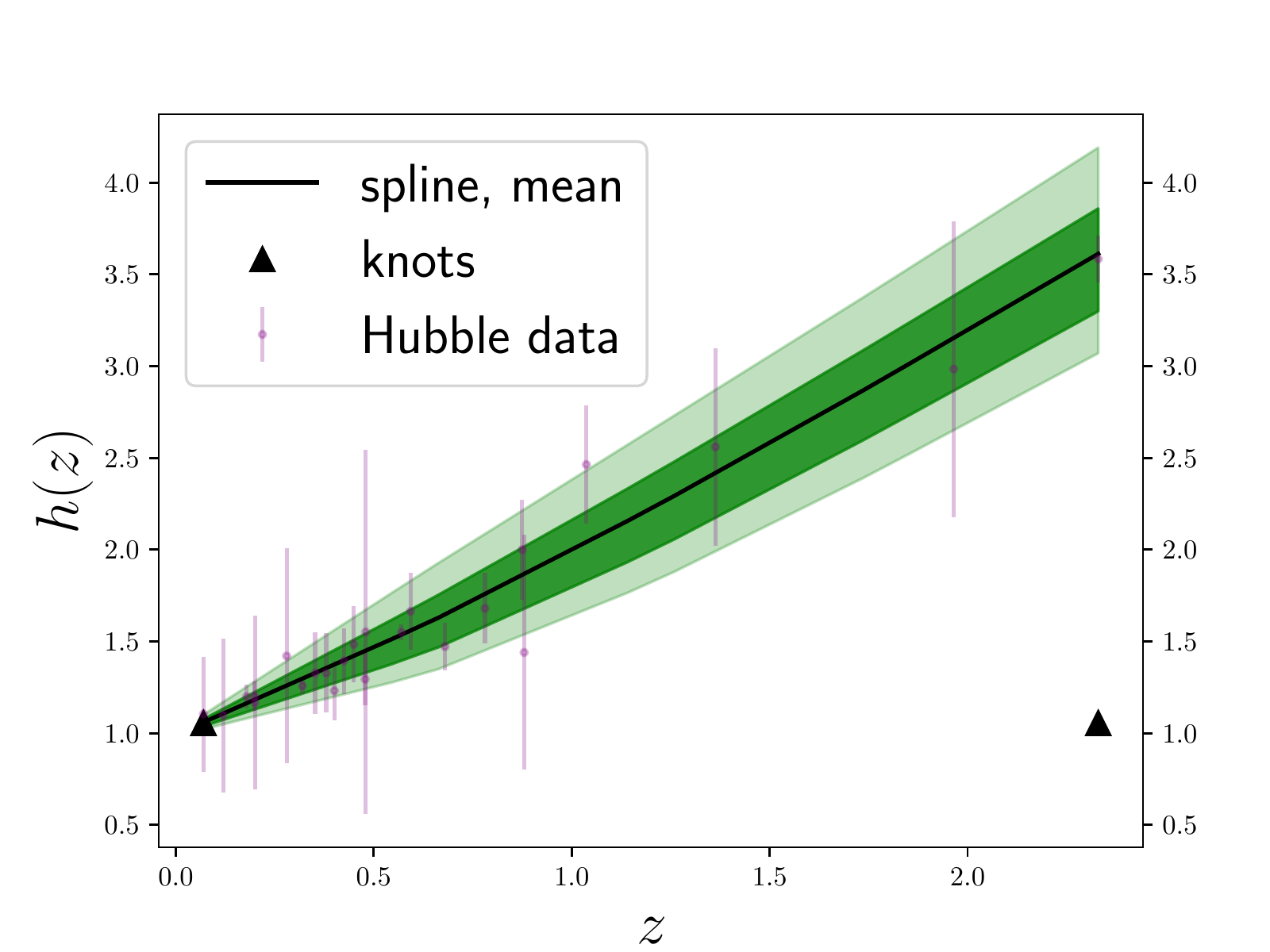}
    \caption{$\chi^2_\textrm{min}/dof=7/20$.}
    \label{fig:hz_mean_hubble_spline_BC03_BAO}
  \end{subfigure}
  ~
  \\

  \begin{subfigure}[b]{0.48\textwidth}
    \includegraphics[width=\textwidth]{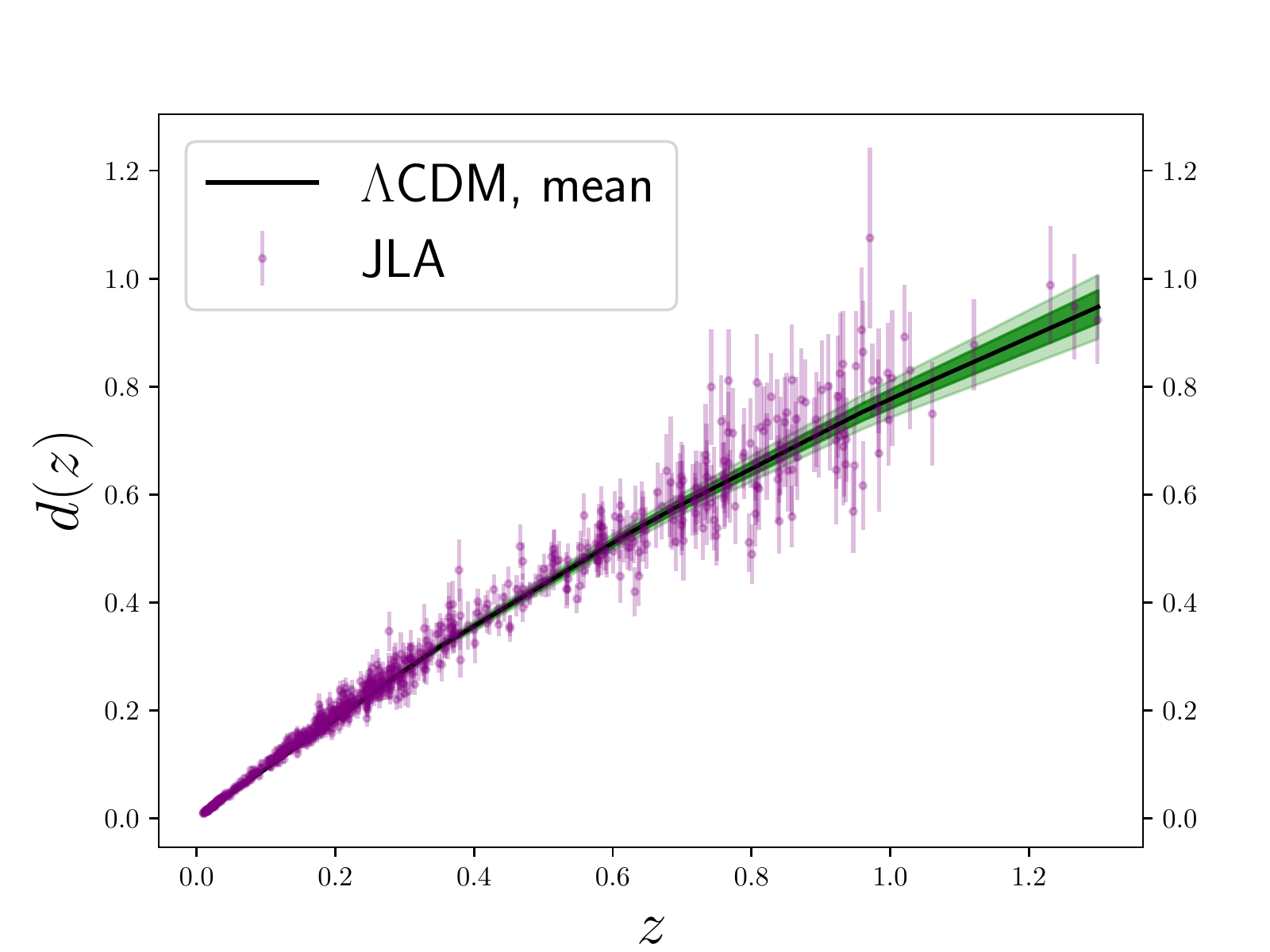}
    \caption{$\chi^2_\textrm{min}/dof=682/732$.}
    \label{fig:d_mean_jla_lcdm}
  \end{subfigure}
  ~
  \begin{subfigure}[b]{0.48\textwidth}
    \includegraphics[width=\textwidth]{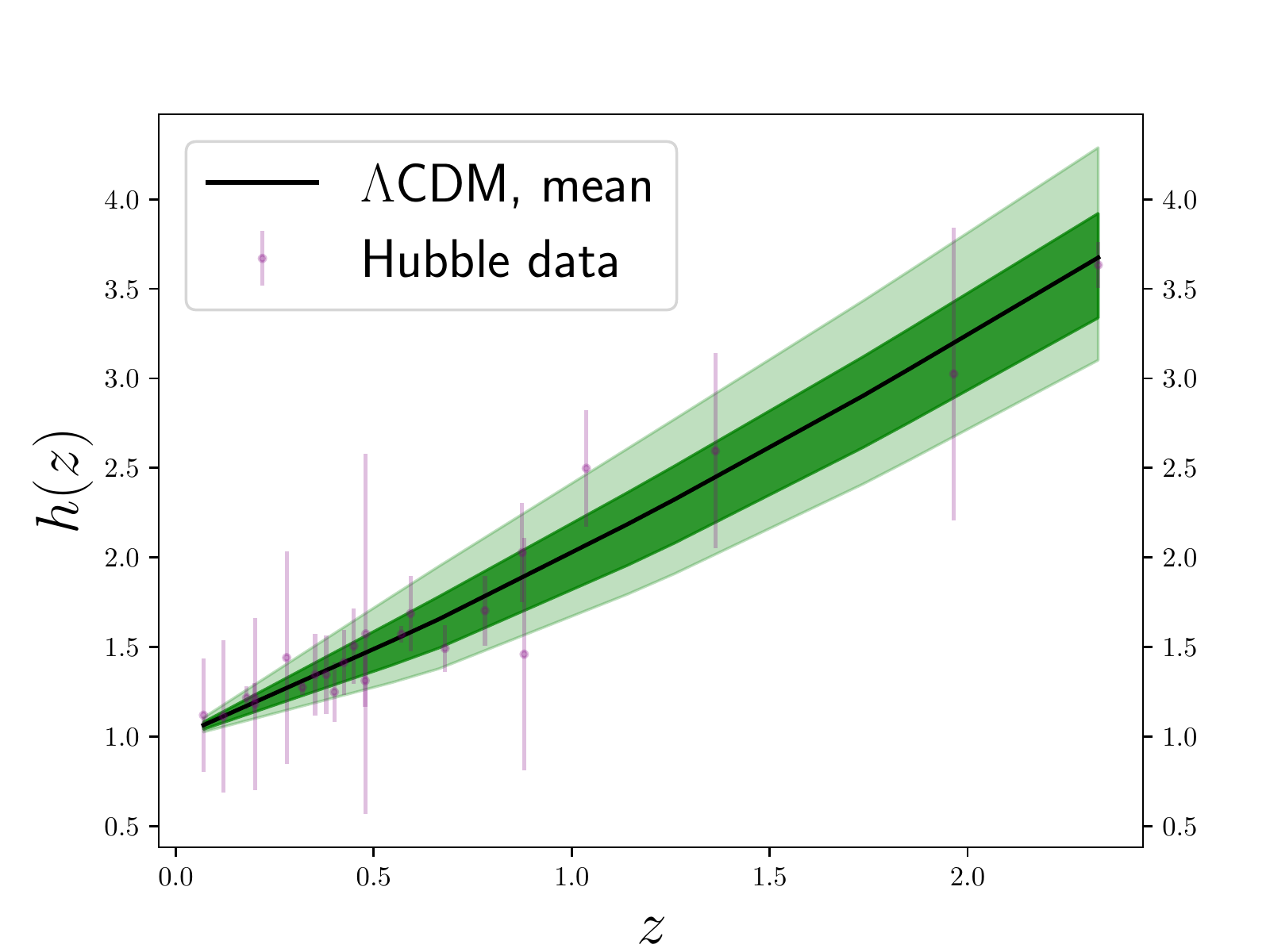}
    \caption{$\chi^2_\textrm{min}/dof=8/20$.}
    \label{fig:hz_mean_hubble_lcdm_BC03_BAO}
  \end{subfigure}
  ~

  \caption{Constraints on the distance (\emph{left}) and the
    expansion rate (\emph{right}) based on the combined JLA and
    BC03+BAO datasets. Black lines show the mean and the contours show
    the marginalised 68\% and 95\% C.I.. We consider polynomial
    (\emph{top}), spline (\emph{middle}) and $\Lambda$CDM
    (\emph{bottom}) fits. Triangles show the location of the spline
    knots.}
  \label{fig:jla_hubble}
\end{figure}

We now move to determining the expansion rate and the consistency condition $k_H(z)$ from the data alone, without backreaction assumptions. In this case, there is no need to look for the maximum of $d_A$, so we consider both polynomials and splines to get a handle on the dependence on the fitting function, and again consider the \LCDM model for comparison. Figure \ref{fig:jla_hubble} shows the distance $d(z)$ for these three cases.  As before, the data points shown in the plots are obtained by fixing the nuisance parameters to their mean values. In the spline case we also show the location of the knots corresponding to the distance values varied in the chain (the knot at $z=0$, fixed by the initial conditions, is not shown).

The three fitting functions give $\chi^2_\textrm{min}/dof$ values that are comparable, though slightly larger for splines. Compared to the polynomial, the spline fit accommodates a wider variety of distances at the same confidence interval (only in small measure due to the fact that marginalisation over the final spline boundary condition involves a further nuisance parameter), and the \LCDM model is of course even more constrained than the polynomial. Otherwise the recovered distances are qualitatively similar, the spline result favouring somewhat longer distances. For the splines, the posterior mean differs from the global best fit by $\sim0.5\sigma$ at lower redshifts. This is caused primarily by degeneracies in parameter space leading to a volume effect with respect to $M_B^1$ in the marginalised constraints, similar to the degeneracies in the backreaction case discussed in appendix \ref{sec:vol_effect}. Mock analysis shows that if $M_B^1$ is fixed to its best fit value, the mean of the distance correctly recovers the expected value. As a result of this degeneracy, the derivatives of the distance are not well recovered. Adding more spline knots does not improve the situation, because in this case the Metropolis--Hastings algorithm is not an efficient enough sampling in the $M^1_B$ direction (alternatives such as Nested Sampling, Hamiltonian/Hybrid Monte Carlo or other techniques could be more appropriate, see \eg \cite{Akeret:2015uha} and references therein).

\subsubsection{Fits to the expansion rate data}

Fits to BC03+BAO Hubble parameter data are shown in figure \ref{fig:jla_hubble}. In this case all fitting functions give similar results. The narrow error contours at small redshifts are partly due to the initial condition $h(0)=1$. The Hubble parameter is better constrained at high redshifts than the backreaction prediction shown in \fig{fig:hz_BestFit_poly_sachs} (note the different redshift ranges of figures \ref{fig:hz_BestFit_poly_sachs} and \ref{fig:jla_hubble}), because in the backreaction case $h(z)$ depends on determination of $\Omn$ using the zero of $d_A'$, which introduces extra spread. Within the errors, the expansion rate does not show any features. It should be noted that our second order polynomial (or two free spline knots) may not have the flexibility to fit features, but the choice reflects the fact that the present data are not very constraining. For results for other fitting methods, see \cite{Busti:2014dua, Verde:2014qea, Bernal:2016gxb}.

Fits to BC03 data are characterised by small $\chi^2/dof \sim 0.4$ due
to large errors. The optimal fit parameters have been tested through
cross-validation to rule out flagrant overfitting. Datapoints obtained
with MaStro stellar modelling give a more reasonable
$\chi^2/dof \sim 0.95$ and are overall consistent with BC03. However,
these goodness-of-fit values should be interpreted with caution. We
expect fits to both BC03 and MaStro to be poorly predictive due to
large errors and small amount of data. For instance, adding just three precise BAO
data points helps to improve the goodness of fit in both cases,
though the overall conclusions are unchanged as the errors are in any
case large.

\begin{table}[t]
  \centering
  \begin{tabular}{ccccc}
    \hline
    \hline
    $H_0$ [km/Mpc/s] & BC03 & BC03+BAO & MaStro & MaStro+BAO \\
    \hline
    \hline
    polynomial & $66.8_{-6.3}^{+6.1}$ & $64.2_{-3.9}^{+5.2}$ & $70.7_{-13}^{+12}$ & $67.7_{-4.8}^{+4.9}$ \\
    spline & $68.8_{-7.1}^{+7.3}$ & $62.5_{-4.6}^{+4.6}$ & $69.0_{-16}^{+15}$ & $68.7_{-5.3}^{+5}$ \\
    \LCDM & $68.4_{-6.3}^{+6.2}$ & $61.7_{-4.5}^{+4.5}$ & $79.6_{-7.5}^{+6.7}$ & $67.7_{-4.8}^{+5.3}$ \\
    \hline
  \end{tabular}
  \caption{The value of the Hubble parameter today $H_0$ for the different fitting functions and datasets.}
  \label{tab:hubble_comp}
\end{table}

In table \ref{tab:hubble_comp} we show the $H_0$ value for different fitting functions and datasets. The values are consistently lower than determinations from nearby SNe Ia (which have mean values from 72.5 to 73.75 km/s/Mpc and 68\% C.I. ranges from 1.7 to 3.2 km/s/Mpc,  depending on the analysis) \cite{Efstathiou:2013via, *Riess:2016jrr, *Cardona:2016ems, Zhang:2017aqn, Follin:2017ljs, Feeney:2017sgx}, although our error bars are large. (The exception is the MaStro only data, which consists of only 8 points, and for which the errors are particularly large.)
This is in agreement with previous determinations of $H_0$ from $H(z)$ data \cite{Busti:2014dua, Verde:2014qea, Busti:2014aoa}.
The differences between fitting functions are within the 68\% C.I. (the effect of different fitting functions was investigated in the context of Gaussian processes in \cite{Seikel:2013fda, Busti:2014aoa}), whereas the differences between stellar evolution models are larger, suggesting caution in the interpretation of the value of $H_0$ derived from cosmic clock data.

The determination of $H_0$ from local SNe is not independent of cosmology, it depends on the value of the deceleration parameter $q_0$ via the series expansion of $D_L(z)$. As discussed in \cite{Feeney:2017sgx}, more positive values of $q_0$ (meaning less acceleration today) give smaller values of $H_0$. If backreaction explains the accelerated expansion, the acceleration is transient and the expansion will at some point start to decelerate \cite{Rasanen:2006zw, *Rasanen:2006kp}. The distance data are consistent with (but, obviously, do not require) considerably less acceleration (or even deceleration) today than in the \LCDM model, as the distance depends on the acceleration via two integrals, and conclusions about $q_0$ depend on the chosen parametrisation \cite{Shapiro:2005nz, *Gong:2006tx, *Elgaroy:2006tp, *Seikel:2007pk, *Seikel:2008ms, *Mortsell:2008yu, *Guimaraes:2009mp, *Serra:2009yp, *Cai:2010qp, *Wang:2010vj, *Park:2010xw, *Pan:2010zh, *Cai:2011px, *Li:2011wb, *Shafieloo:2012ht, *Aviles:2012ay, *Magana:2014voa, *Qing-Guo:2016ykt, *Ade:2015rim, Cattoen:2007id, *Visser:2009zs}. For the BC03+BAO expansion rate data, deceleration today is within the 68\% limits for the splines, and well within the 95\% contours for the polynomials. However, this is mostly indicative of the poor constraining power of the $H(z)$ data, as deceleration today is also within the 68\% contours of the \LCDM model. (For MaStro+BAO, deceleration today is within the 95\% contours in all of the three cases.)

The effect of late-time deceleration on the distance can be compensated by having more acceleration in the past. Indeed, if backreaction explains the accelerated expansion, the acceleration can naturally be stronger than in the \LCDM case, with an effective equation of state more negative than $-1$, being preceded by stronger deceleration \cite{Rasanen:2006zw, *Rasanen:2006kp, Boehm:2013qqa}. A period with extra deceleration corresponds to effective negative energy density, which is unnatural in the FRW framework, but expected in the backreaction case. It has been noted that there is tension between CMB and Ly$\a$ BAO data at redshift $z=2.34$, which is difficult to explain in FRW models, but can be solved by negative energy density \cite{Delubac:2014aqe, Aubourg:2014yra, Ade:2015xua}. This is around the redshift range where one could expect extra deceleration from backreaction \cite{Boehm:2013qqa}, but not much can be read into such possible hints without a backreaction prediction for $h(z)$, whether from theoretical calculations or more precise distance data and the relation \re{hbr} between $d(z)$ and $h(z)$.

\subsubsection{Consistency condition $k_H(z)$} \label{sec:kHdata}

\begin{figure}[t]
  \centering

  \begin{subfigure}[b]{0.48\textwidth}
    \includegraphics[width=\textwidth]{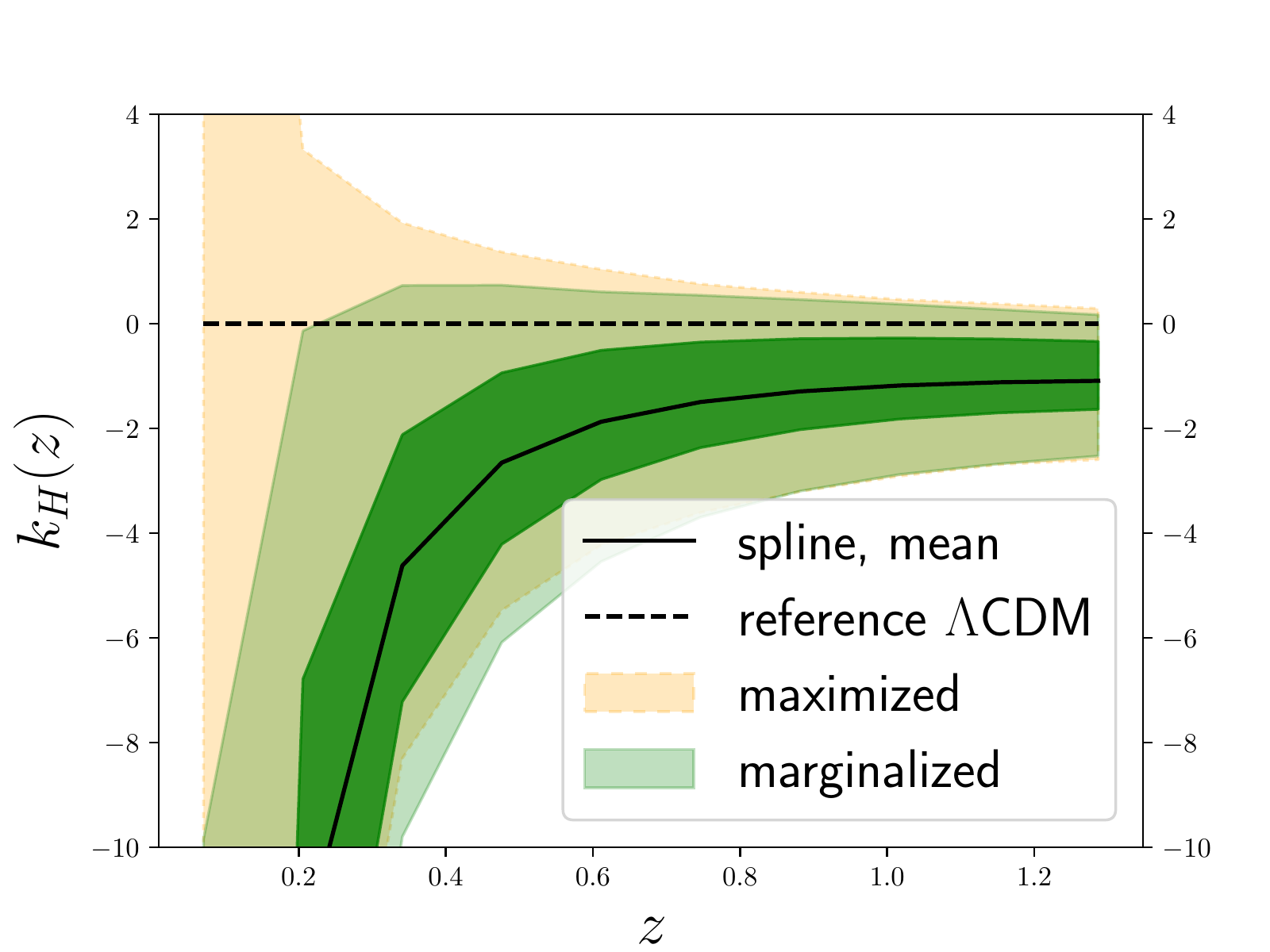}
\caption{}
    \label{fig:kH_mean_jla_hubble_spline_BC03_BAO}
  \end{subfigure}
  ~
  \begin{subfigure}[b]{0.48\textwidth}
    \includegraphics[width=\textwidth]{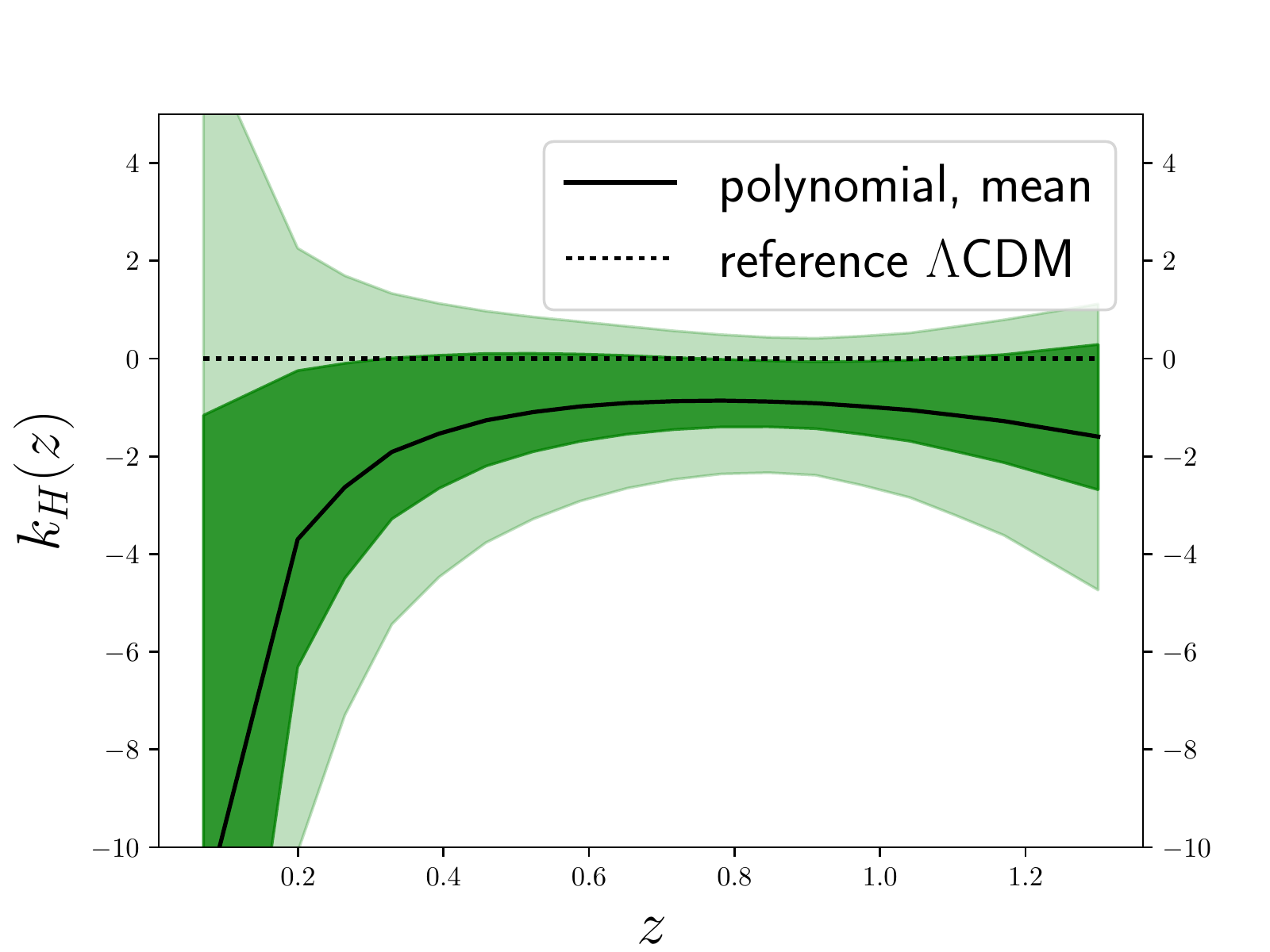}
    \caption{}
\label{fig:kH_mean_jla_hubble_poly_BC03_BAO}
  \end{subfigure}

  \caption{The consistency condition $k_H(z)$ determined from JLA and
    BC03+BAO data. We consider the spline (\emph{left}) and polynomial
    (\emph{right}) fits. The marginalised contours (green) show the
    68\% and 95\% C.I.. In the spline case, we also plot the
    maximised 95\% C.I. (orange).} \label{fig:jla_hubble_kH}
\end{figure}

In figure~\ref{fig:jla_hubble_kH} we show the consistency condition function $k_H(z)$ obtained by fitting to the JLA+BC03+BAO data using both splines and polynomials. The polynomial fit shows no violation of the FRW consistency condition. In the spline case, there appears to be a significant deviation from $k_H=0$, but this has to be treated with caution. As explained in section \ref{sec:dhdata_dist} (and discussed further in appendix \sec{sec:vol_effect}), in the spline analysis we can have strong shifts in the 1-dimensional projected statistics due to the volume effect introduced by $M_B^1$. The fact that the maximised contours (also plotted in in figure~\ref{fig:jla_hubble_kH}) do not show significant deviations from the FRW case is consistent with the interpretation that this is projection effect. When fitting JLA+MaStro+BAO data with splines, the marginalised 95\% contours are consistent with $k_H=0$. The large errors at low redshift are due to the fact that the denominator of $k_H$ goes to zero as $z^2$ when $z\to0$. The numerator diverges at least as rapidly, so the result is not divergent, but errors cause a mismatch between the measured values of the denominator and numerator. At large redshifts, in turn, the distance is less well determined. The best constraints over the redshift range are $-2.32<k_H<0.40$ at $z=0.9$ for the polynomials and $-2.53<k_H<0.17$ at $z=1.3$ for the splines. While polynomials are better constrained at intermediate redshifts, splines give smaller errors at large $z$, reaching similar errors on $k_H$ as polynomials despite the fact that the distance is less well determined. This is not surprising since in the spline case we also fit the first derivative $d_L'(z_n)$ at the last spline knot $z_n$, providing a better determination of $d'(z)$  in \re{kH} than the polynomial case.

We also fit the polynomial $d(z)$ to the JLA+BC03+BAO data assuming a constant $k_H$. We use the FRW relation $h^2=(1-k_H d^2)/d'^2$ with a constant $k_H$. The result is sensitive to the assumed value for $H_0$, as found in \cite{Wei:2016xti}. Taking $H_0=70$ km/s/Mpc we get $k_H=-0.19_{-0.25}^{+0.29}$, whereas $H_0=67$ km/s/Mpc and $H_0=73$ km/s/Mpc give, respectively, $k_H=-0.55_{-0.29}^{+0.33}$ and $k_H=0.06_{-0.23}^{+0.24}$. (The results for the JLA+MaStro+BAO combination are similar, within the error bars.) For comparison, fitting the \LCDM model to the JLA data alone gives the (by construction constant) value $k=-\Omega_{K0}=-0.23_{-0.27}^{+0.25}$. (The BC03+BAO data gives $k=-\Omega_{K0}=-0.65_{-0.34}^{+0.46}$.) Taking into account the dependence on the parametrisation, on the debated value of $H_0$ and on the different stellar evolution models, the constraints can be conservatively summarised as $|k_H|\lesssim1$. If the universe is well described by the FRW model, these numbers are a direct constraint on the spatial curvature parameter.
A combination of the local $H_0$ value and the distance to the last scattering surface, which can be determined model-independently from the CMB \cite{Vonlanthen:2010cd, *Audren:2013nwa, Audren:2012wb}, gives the constraint $k_H(z=1090)<0.1$ from the requirement that the universe is large enough to contain the last scattering surface \cite{Rasanen:2014mca}.

These observational constraints on a constant $k_H$ are of the same order as the backreaction prediction shown in \fig{fig:kH_BestFit_jla_poly_sachs}. However, in the case when $k_H$ is allowed to vary they loosen considerably; even at the best constrained redshift, $z \approx 1$ for polynomials, our constraint on $k_H(z)$ is about a factor of 3 weaker than the bound on a constant $k_H$. Observational analyses where $k_H$ is taken to be constant therefore cannot be directly used as constraints on the magnitude of $k_H(z)$ that varies with $z$.

\begin{figure}[t]
  \centering

  \includegraphics[width=0.7\textwidth]{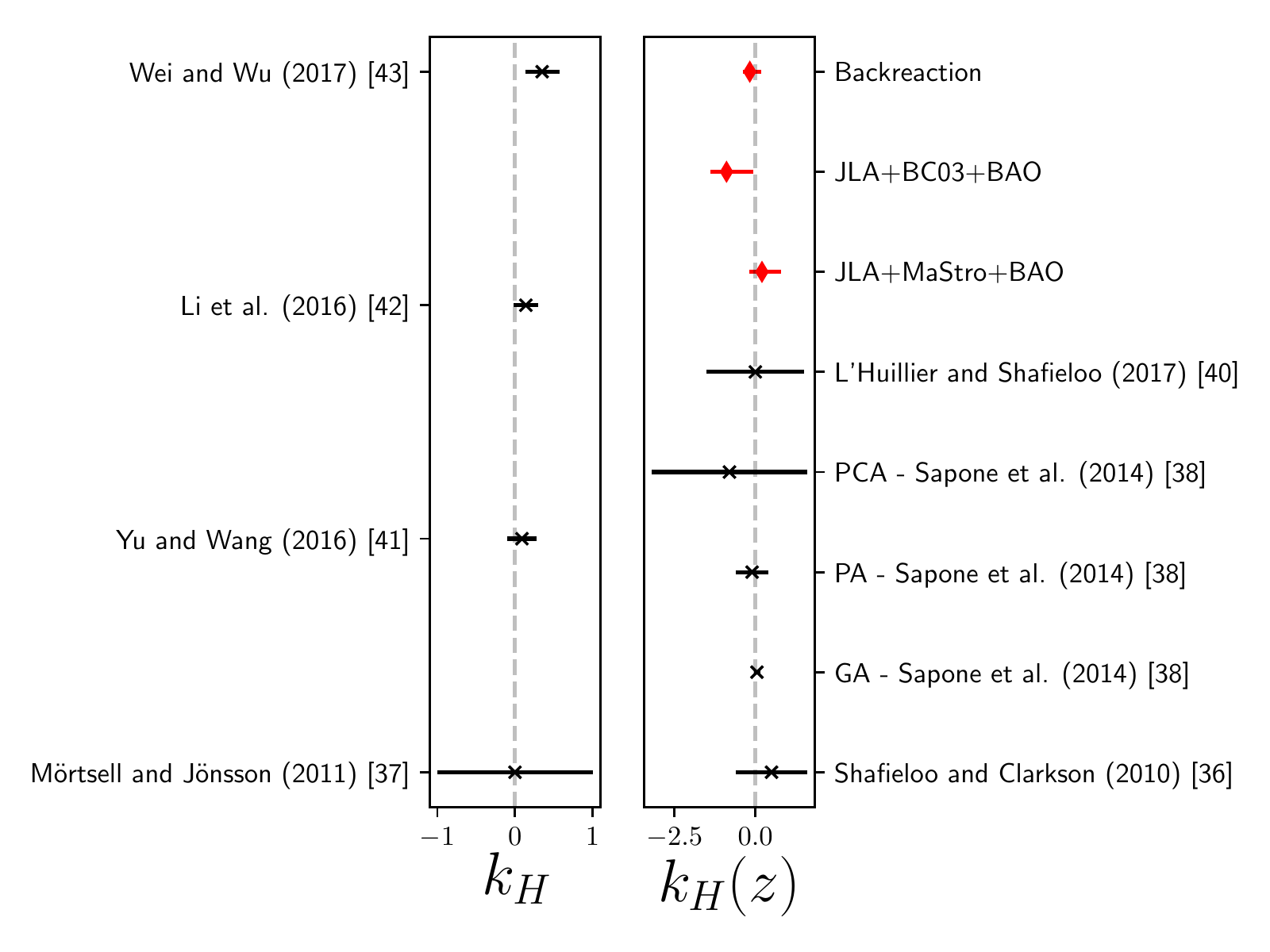}

  \caption{Comparison of 68\% C.I. constraints on the parameter $k_H$ when it is
  constant (\emph{left}) and redshift-dependent (\emph{right}). In the latter case we consider the tightest $k_H(z)$
    determination over the redshift range considered in the
    respective analysis. PCA stands for principal component analysis, PA for Pad\'e approximants
    and GA for genetic algorithms.
    Our results for a redshift-varying $k_H(z)$ are shown as
    red diamonds.}
  \label{fig:kH_literature}
\end{figure}

Our results roughly agree with those in the literature, with the caveat that the constraints depend on the dataset and the fitting method. Results for a constant $k_H$ are $|k_H|\lesssim1$ (95\% C.I.) \cite{Mortsell:2011yk}, $k_H=0.09\pm0.19$ (using a method applicable only for $|k_H|\ll1$) \cite{Yu:2016gmd}, $k_H=0.35\pm0.22$ \cite{Wei:2016xti} and $k_H=0.14\pm0.16$ \cite{Li:2016wjm}. Due to different model-dependence of the datasets and varying methods of fitting and determining the error bars, the strongest quoted constraints are not necessarily the most reliable. These limits are summarised in \fig{fig:kH_literature}.

In figure \ref{fig:kH_literature} we also show results in the literature in the case when redshift dependence is taken into account. The results in the literature are $k_H(z)=0.5\pm1.1$ \cite{Shafieloo:2009hi}, $k_H(z)=-0.8\pm2.4$ (with principal component analysis, PCA), $k_H(z)=-0.1\pm0.5$ (with Pad\'e approximants, PA), $k_H(z)=0.05\pm0.1$ (with genetic algorithms, GA) \cite{Sapone:2014nna} and $k_H(z)=0\pm1.5$ \cite{LHuillier:2016mtc}. (The error bars from GA are very small, but they have been estimated with bootstrap methods \cite{Nesseris:2010ep}.)  These are limits for the redshifts for which the constraints are the tightest, determined by the balance between the low redshift mismatch divergence and poor accuracy of high redshift data. Apart from the case of genetic algorithms, the constraints loosen considerably. In \cite{Cai:2015pia, LHuillier:2016mtc} it was also checked whether the data are consistent with the spatially flat FRW relation $hd'=1$, but the error bars cannot be readily interpreted in terms of $k_H$.

\subsection{Forecast} \label{sec:forecasts}

\begin{figure}[t]
  \centering

  \begin{subfigure}[b]{0.48\textwidth}
    \includegraphics[width=\textwidth]{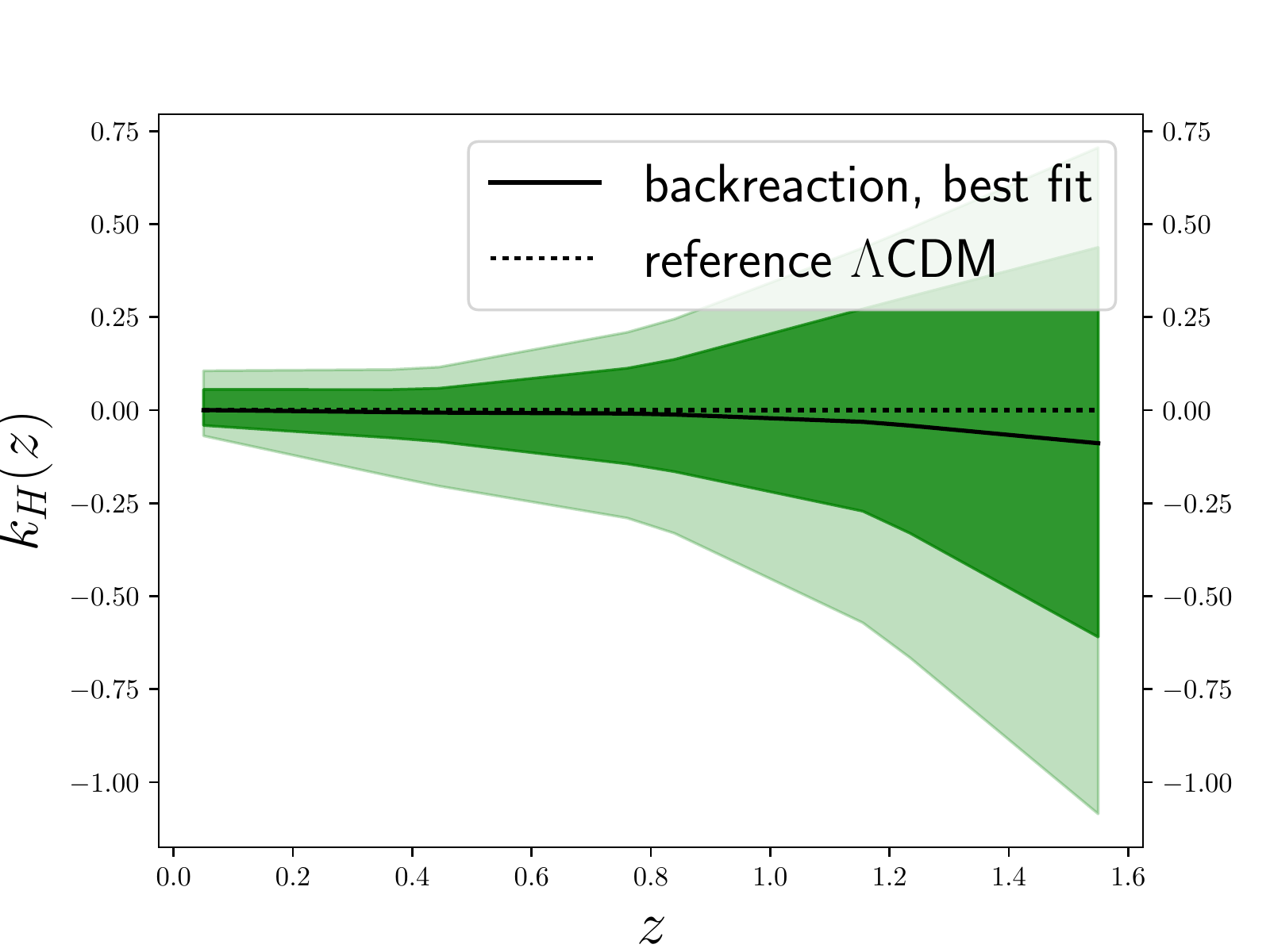}
    \caption{}
  \end{subfigure}
  ~
  \begin{subfigure}[b]{0.48\textwidth}
    \includegraphics[width=\textwidth]{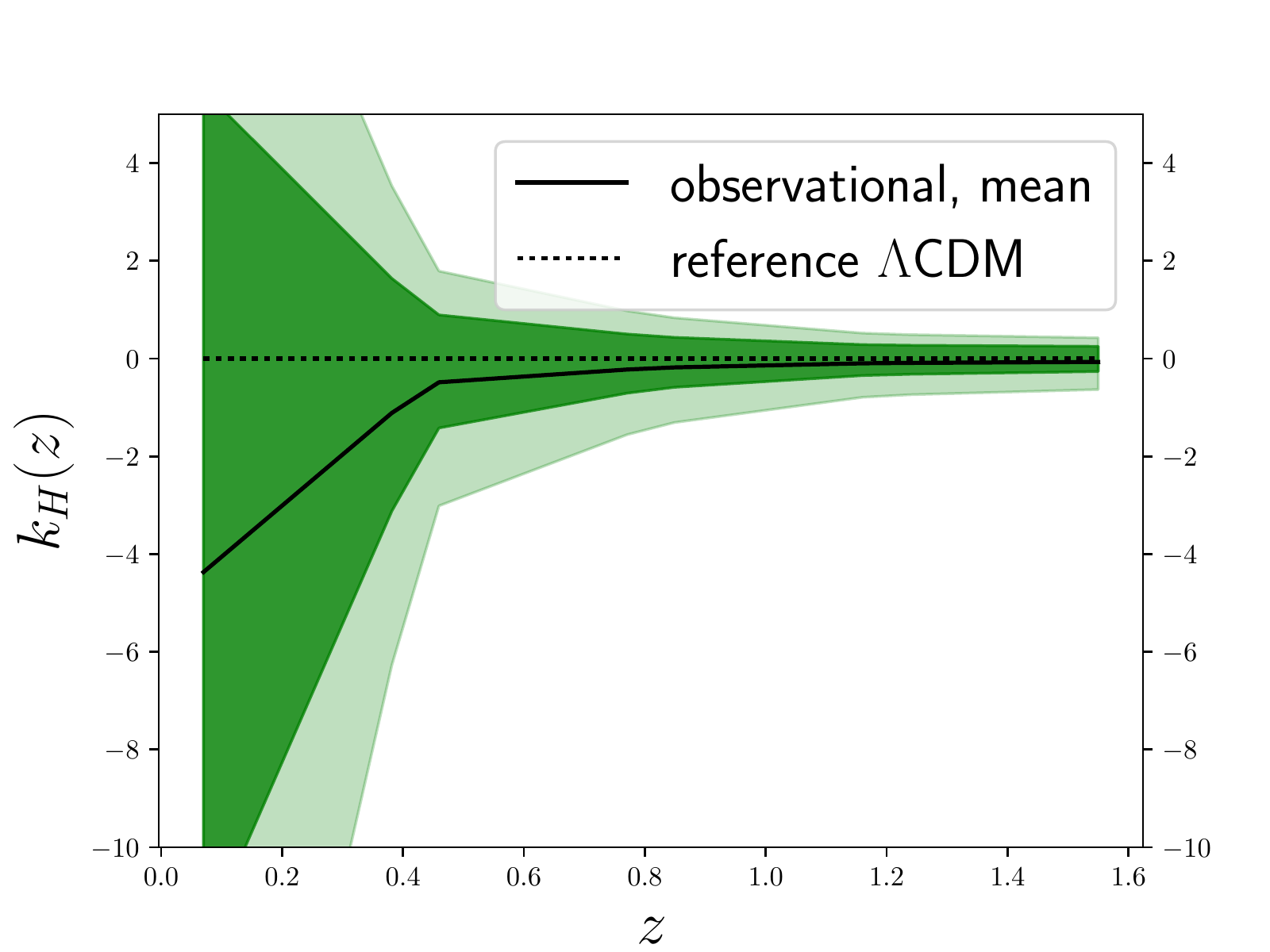}
    \caption{}
  \end{subfigure}

  \caption{Forecast for the consistency condition $k_H(z)$ based on a
    polynomial fit to a future SNIa survey loosely modelled on
    LSST+Euclid, and to current expansion rate data supplemented by future Euclid data.
    We show the backreaction prediction based on SNe Ia (\emph{left}), and the purely observational result based on SNIa and expansion rate data (\emph{right}).}
\label{fig:forecasts}
\end{figure}

Finally, let us forecast how well next generation experiments can constrain $k_H(z)$.

We consider $10^5$ logarithmically spaced SNe Ia in the range $0.05<z<1.55$. This is loosely modelled on a future catalogue combining observations from experiments such as LSST\footnote{\url{https://www.lsst.org/}} and Euclid\footnote{\url{https://www.euclid-ec.org/}} \cite{Astier:2014swa}. We assume fractional errors on the distance modulus equal to the mean of JLA fractional errors $\sigma_{\mu}/\mu = 4.5 \times 10^{-3}$ (the dispersion in the fractional error for different data points, about $10^{-3}$ in JLA, is negligible for our purposes). We neglect nuisance parameters, as we only aim at a rough estimate. Given the larger redshift range spanned compared to JLA, fourth order polynomials do not recover the data well enough, so we model the luminosity distance with a fifth order polynomial.

Euclid \cite{Laureijs:2011gra} is also expected to measure differential ages of passive galaxies within the range $1.5\lesssim z\lesssim2$. (See also \cite{Zhang:2010ic, Carvalho:2011qw, Sapone:2014nna} for discussion of future $H(z)$ data.) We add the five simulated Hubble parameter datapoints from figure 2 of \cite{Moresco:2015cya} to the BC03 data points listed in table \ref{tab:H}. We consider the same second-order polynomial fit as before and marginalise over $H_0$.

The data are rescaled to a fiducial flat $\Lambda$CDM cosmology
with $\Omn=0.3$ and $H_0=70$ km/s/Mpc (we neglect realisation noise).

Given the different polynomial order of the SNIa fit, constraints on the distance coefficients cannot be readily compared to JLA results. However, the projected Hubble parameter results can be directly compared to those based on a BC03+BAO catalogue, rescaled at the fiducial cosmology. Error contours are similar, indicating that replacing the accurate BAO datapoints by the model-independent Euclid differential age datapoints will not degrade error bars even though the simulated Euclid errors on $H(z)$ are about $\sigma_H=30$ km/s/Mpc, contrasted with the BAO errors of $\sigma_H=3\ldots8$ km/s/Mpc.

On the left panel of \fig{fig:forecasts} we show constraints on the consistency condition $k_H(z)$ based on future SNIa data alone, calculating $h(z)$ from the backreaction relation \re{hbr}.  As we assume the existence of a maximum in the angular diameter distance, we show the maximised (rather than marginalised) statistics, as before. Compared to \fig{fig:kH_BestFit_jla_poly_sachs}, the forecasts show an improvement of a factor of 6 at low redshifts, with the tightest constraint at the minimum redshift $z=0.05$ having the 68\% C.I. range $-0.04<k_H<0.06$ and 95\% C.I. range $-0.07<k_H<0.11$. Since in the forecast we keep nuisance parameters fixed, this improvement has to be interpreted as a lower possibly reachable bound on the errors.
On larger redshifts the errors are comparable to present data. This indicates that, at large redshifts, the limiting systematic of this method is the bias introduced when determining $\Omn$ (see \sec{sec:bias}). Note that the upper limit of our projected SNIa data are close to the redshift of the angular diameter distance maximum in our fiducial model; the precision may vary depending on the real position of the maximum and reach of the SNIa data.

On the right panel of \fig{fig:forecasts} we show the joint marginalised constraints on the consistency condition $k_H(z)$ based on future SNIa and Hubble parameter data. Compared to \fig{fig:jla_hubble_kH}, error contours decrease significantly, especially at large redshifts. The improvement is mainly driven by the larger number of SNIae. The tightest 68\% C.I. limit is $-0.26<k_H<0.25$ (the 95\% C.I. limit is $-0.63<k_H<0.43$, reached at the largest redshift $z=1.55$. This is a factor of 3 improvement compared to current data.
This is comparable to the 68\% C.I. limit $|k_H(z)|\lesssim0.2$ obtained in \cite{Sapone:2014nna} and based on a similar future SNIa catalogue as the one we have considered, but including Euclid BAO Hubble parameter data.  Given that the expected magnitude from backreaction is $|k_H(z)|\sim0.1\ldots1$ \cite{Boehm:2013qqa}, future observations are expected to probe the theoretically interesting region, but not cover it exhaustively.

\section{Conclusions} \label{sec:conc}

We have fitted the JLA SNIa distance data with a fourth order polynomial, determined the expansion rate as a function of redshift using the backreaction relation \re{hbr} and calculated the resulting backreaction prediction for the FRW consistency functions $k_H(z)$, $k_S(z_\ml, z_\ms)$ and $k_P(z)$.  This method of determining the expansion rate requires the angular diameter distance to have a maximum. As the maximum is typically slightly outside the JLA data range (or is not clearly visible in the data), this means that many good fits have to be discarded. This selection effect leads to significant bias in marginalised parameters, but the maximised statistics give reliable results.  The best fits for all three functions are approximately constant and equal to $-0.2$. The 95\% C.I. range for $k_H$ and $k_P$ (which are essentially indistinguishable) varies between $-0.7<k_H,k_P<0.4$ at small redshift and $-0.6<k_H,k_P<1.3$ at $z=1.3$, and $k_S(z_\ml, z_\ms)$ is of the same order of magnitude.

We have also determined the function $k_H(z)$ directly without cosmological assumptions (except for those associated with BAO data reduction) by combining the JLA SNIa distance data with cosmic clock and BAO expansion rate data. We have compared the results of fitting with polynomials or splines, and using either the BC03 or MaStro stellar population evolution models for cosmic clocks. We have carefully checked for bias and the reliability of the error bars with mock data and, in the case of splines, further guarded against overfitting with a cross-validation analysis. For splines we see a similar misleading volume effect as in the backreaction case, highlighting the importance of validating methods with mock data.
The details show significant dependence on the stellar evolution model and the fitting method, but the overall trends are similar. At $z\lesssim0.4$, errors are overwhelming due to the fact that $k_H$ is a ratio between two terms that vanish at $z=0$ and do not match precisely due to errors. The best 95\% C.I. constraint, $-2.32<k_H<0.40$, is at $z=0.9$ for the polynomial fit (a similar constraint holds for splines) to the JLA+BC03+BAO data. Replacing BC03 data by MaStro data points gives  $-0.86<k_H<1.13$ at $z=0.8$. In comparison, the constraints for a constant $k_H$ are nearly 3 times stronger, demonstrating that limits derived for a constant $k_H$ in the literature cannot be directly applied to backreaction, which (if significant) is expected to produce a $k_H$ with significant $z$-variation. Furthermore, non-trivial redshift correlations and highly asymmetric error contours show the importance of consistently modelling the covariance matrix and the full non-Gaussian information, suggesting that care should be taken when using fitting methods such as Gaussian processes that rely on assumptions about correlations between different redshifts.

We have considered the value of $H_0$ determined from the $H(z)$ data, and found that while the dependence on the fitting function is well within the 1$\sigma$ errors, the effect of the adopted stellar evolution model is larger, suggesting caution in the interpretation of these values, which are smaller than those determined from local SNe, although with large error bars.

Finally, we have done a forecast of the improvement expected from a future SNIa survey loosely modelled on LSST and Euclid and additional cosmic clock $H(z)$ datapoints from Euclid galaxy differential age measurements. Observational constraints on $k_H(z)$ tighten by up to a factor of 6 for the backreaction case and 3 for the model-independent case, reaching the order of magnitude $|k_H|\sim0.1$ where signatures of backreaction are expected if it is significant. However, a $z$-dependent $k_H$ of this order of magnitude cannot be ruled out by next generation data of the kind we consider. The accuracy may be improved if the maximum of the angular diameter distances will be clearly determined by the data, which depends on a combination of redshift coverage and errors.  Note also that our forecast is based on the spatially flat \LCDM model. The result may be different if the expansion rate or the distance has features, as is expected if backreaction is significant. This is difficult to take into account, as there is no reliable prediction for the change in the expansion rate due to backreaction, and the current data does not have strong constraining power for features. Nevertheless, we conclude that upcoming observations are expected to probe an interesting range of $k_H$ and that general model-independent tests such as the FRW consistency conditions continue to complement more precise model studies.

\acknowledgments

We thank Elisabetta Majerotto for contributing at an early stage of this work, and Martin Kunz and Jussi V\"{a}liviita for useful discussions.

\appendix

\section{Roughness parameter for the spline fit} \label{sec:rough}

We do a cross-validation analysis in the spline case to avoid overfitting. Instead of computing the likelihood based directly on $\chi_{SN}^2$ as in \eqref{eq:chi2_jla}, we consider the following effective $\chi^2$:
\begin{eqnarray} \label{eq:smooth_spline}
  \chi^2_{eff}[d_L] &=& \sum_{i,j=1}^{N} \left\{{\hat
                        \mu}[d_L(z_i)]-\mu[d_L(z_i)]\right\}
                        \mathrm{C}^{-1}(z_i,z_j) \left\{{\hat
                        \mu}[d_L(z_j)]-\mu[d_L(z_j)]\right\}
                        \nonumber \\
                    &&+ \lambda \int_{\ln z_0}^{\ln z_N}
                       \left[d_L''(\ln z)\right]^2 d\ln z
                       \nonumber \\
                    &=& \chi_{SN}^2 + \lambda \int_{\ln z_0}^{\ln z_N}
                        \left[d_L''(\ln z)\right]^2 d\ln z \;,
\end{eqnarray}
where the constant $\lambda \geq 0$ is a roughness parameter. The
first term is the $\chi_{SN}^2$ introduced in
\eqref{eq:chi2_jla}. The explicit sum over the $N$ data points at
redshifts $z_i$ highlights that in general they differ from the spline
knots $z_k$, where $k=0,\ldots,1+n$.  The form of the penalty factor
is inspired by smoothing spline algorithms
\cite{green1993nonparametric,Sealfon:2005em,Peiris:2009wp}. The
integral multiplying $\lambda$ takes into account the mean curvature
of the spline (the integrand is squared since we are not
interested in the sign of the curvature).  Values $\lambda>0$ penalise
irregularly oscillating functions that may fit noise (in the limit
$\lambda\to\infty$ only linear functions would be allowed).

We select the optimal value of $\lambda$ by demanding that if the
underlying function is correctly recovered, it should
accurately predict new independent data. This requirement is implemented by
performing a $2$-fold cross-validation (CV) \cite{Sealfon:2005em,
  Peiris:2009wp, Pedregosa:2012toh}.  First, we fix a value for $\lambda$
and divide the data into two halves, $A$ and $B$, homogeneously
distributed in redshift (in practice, $A$ is given by the odd
rows of a given catalogue and $B$ by the even rows). We then perform MCMC
minimisation of $\chi^2_{eff}[d_L]$ on the first half $A$
that serves as a training dataset, determining the best fit spline
parameters. The $\chi^2_{eff}$ of the second half $B$, which serves as
a test dataset, is then computed at those best fit values, giving the score
CV$_{AB}$.  If the training dataset is overfitted, the model so
determined will predict poorly the test dataset, resulting in a large
score CV$_{AB}$. The roles of the two halves are switched and the
procedure is repeated, determining the score CV$_{BA}$. The total
score CV$_{AB}$+CV$_{BA}$ is stored for the parameter $\lambda$.
The optimal $\lambda$ value is the one that minimises
CV$_{AB}$+CV$_{BA}$, reducing the risk of overfitting.

Given that at each step full MCMC chains are required, in practice we
choose the optimal one among the five possibilities $\lambda=0, 0.01, 0.1, 1,
10$. The second derivatives appearing in (\ref{eq:smooth_spline})
are computed as linear interpolations of the spline algorithm
derivatives at the knot values. In our cubic spline algorithm
we require second derivatives to be continuous and to vary linearly
between two knots. Since we use second derivatives only to set the
roughness parameters, the fact that they may be a poor approximation
is not a concern.

The cross-validation analysis suggests that the selected number of spline knots does not overfit the supernova nor the Hubble parameter data, as they prefer the values $\lambda=0$ and $\lambda=0.01$, respectively (and in the Hubble parameter case, the CV score of the $\lambda=0$ case is very similar to the one for $\l=0.01$).

\section{Redshift correlations} \label{sec:red_corr}

\begin{figure}[t]
  \centering

  \begin{subfigure}[b]{0.48\textwidth}
    \includegraphics[width=\textwidth]{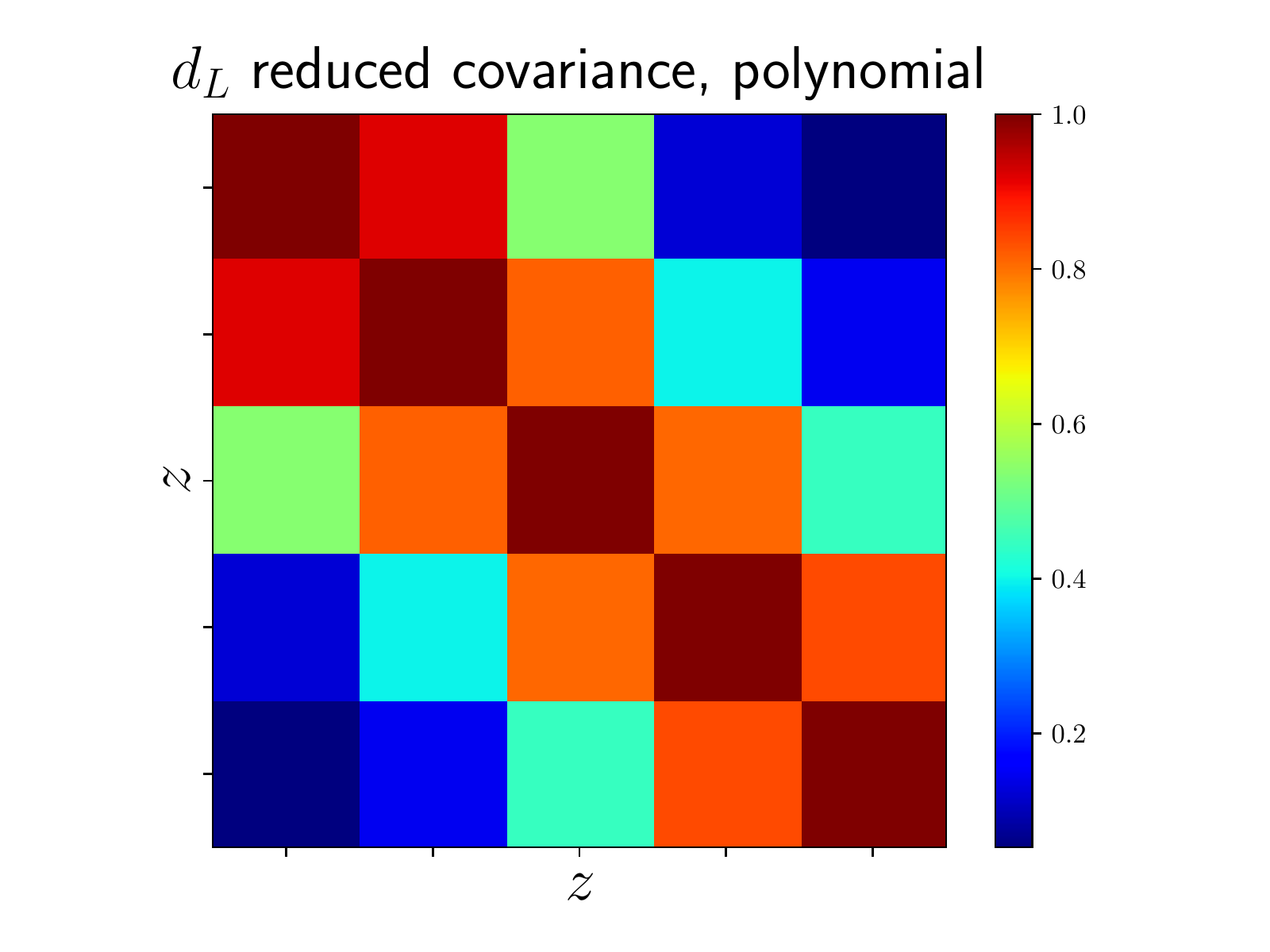}
    \caption{}
    \label{fig:dL_cov_jla_poly_sachs}
  \end{subfigure}
  ~
  \begin{subfigure}[b]{0.48\textwidth}
    \includegraphics[width=\textwidth]{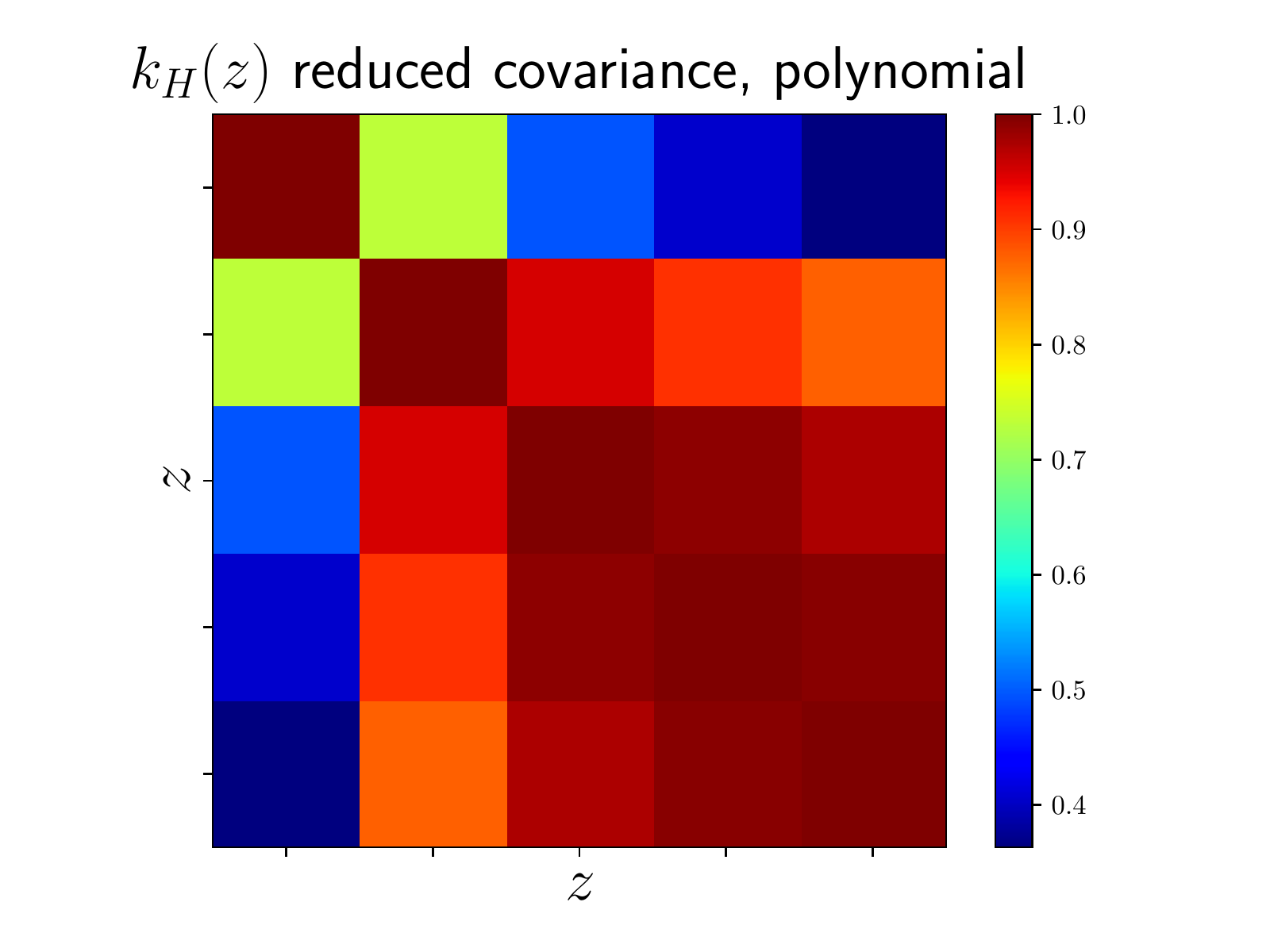}
    \caption{}
    \label{fig:kH_cov_jla_poly_sachs}
  \end{subfigure}
  \\

  \caption{Reduced correlation matrices related to the backreaction
    fits shown in figure~\ref{fig:jla_sachs} for the luminosity
    distance (\emph{left}) and the consistency condition $k_H$
    (\emph{right}). Each pixel corresponds to one of the five linearly
    spaced redshifts within JLA data range $0.01<z<1.3$ where the
    respective function has been constrained.}
  \label{fig:jla_sachs_cov}
\end{figure}

As discussed in section \ref{sec:der}, our MCMC-based approach
provides error contours on each function of redshift $f_i = f(z_i)$
without restrictive assumptions about the underlying probability
distribution. Furthermore, the covariance matrix $\textbf{C}_f$
computed from the chains provides information not only on the variance
$\sigma_{f_i}$, but also on the correlations among the functions
$f_i$, $f_j$ at different redshifts $z_i$, $z_j$. To visualise to
correlation between different redshifts we introduce the reduced
correlation matrix
\begin{equation}
  {\bf r} =
  \left[{\rm diag}(\textbf{C}_f)\right]^{-1/2} \textbf{C}_f
  \left[{\rm diag}(\textbf{C}_f)\right]^{-1/2}
  \;,
\end{equation}
where ${\rm diag}(\textbf{C}_f)$ is the matrix of the diagonal
elements of a given covariance matrix $\textbf{C}_f$.\footnote{The
  elements $r_{ij}$ of the reduced correlation matrix take values
  $-1<r_{ij}<1$, and are equal to 1 on the diagonal $i=j$ by
  construction.  If $|r_{ij}| \ll 1$, no significant correlation is
  present between the corresponding $i$-th and $j$-th parameters
  (\eg between the values of $k_H$ at redshifts $z_i$ and
  $z_j$).}

Figure \ref{fig:jla_sachs_cov} shows the reduced correlation matrix
for the luminosity distance $d_L(z)$ and the consistency function $k_H(z)$ for
the polynomial fit to JLA data when $h(z)$ is determined from the backreaction relation \re{hbr}.
In both cases there are important off-diagonal contributions. Redshift correlations are
particularly significant for the consistency
relation, partially because the Hubble parameter is obtained by integrating the distance over the redshift. While this
correlates  different redshifts non-trivially, the information is
easily propagated through the MCMC algorithm. Besides these two
examples, we have verified that all the functions considered in our
analysis have important redshift correlations. The correlations in the $\Lambda$CDM fit are similar as in the polynomial case, while in the spline case the off-diagonal terms are even larger.

Reconstruction techniques such as Gaussian processes rely on ansatzes
about redshift correlations (see \cite{Seikel:2013fda, Busti:2014aoa} for
discussion of the effects of different covariance functions). It is
then important to model such non-trivial
correlations correctly. Furthermore, as the strongly asymmetric error contours
in the main text show, not all the functions can be well described by
their second-order statistics alone and it is necessary to be able to
model fully non-Gaussian profiles.

\section{Volume effect due to marginalisation} \label{sec:vol_effect}

\begin{figure}[t]
  \centering
  \includegraphics[width=0.5\textwidth]{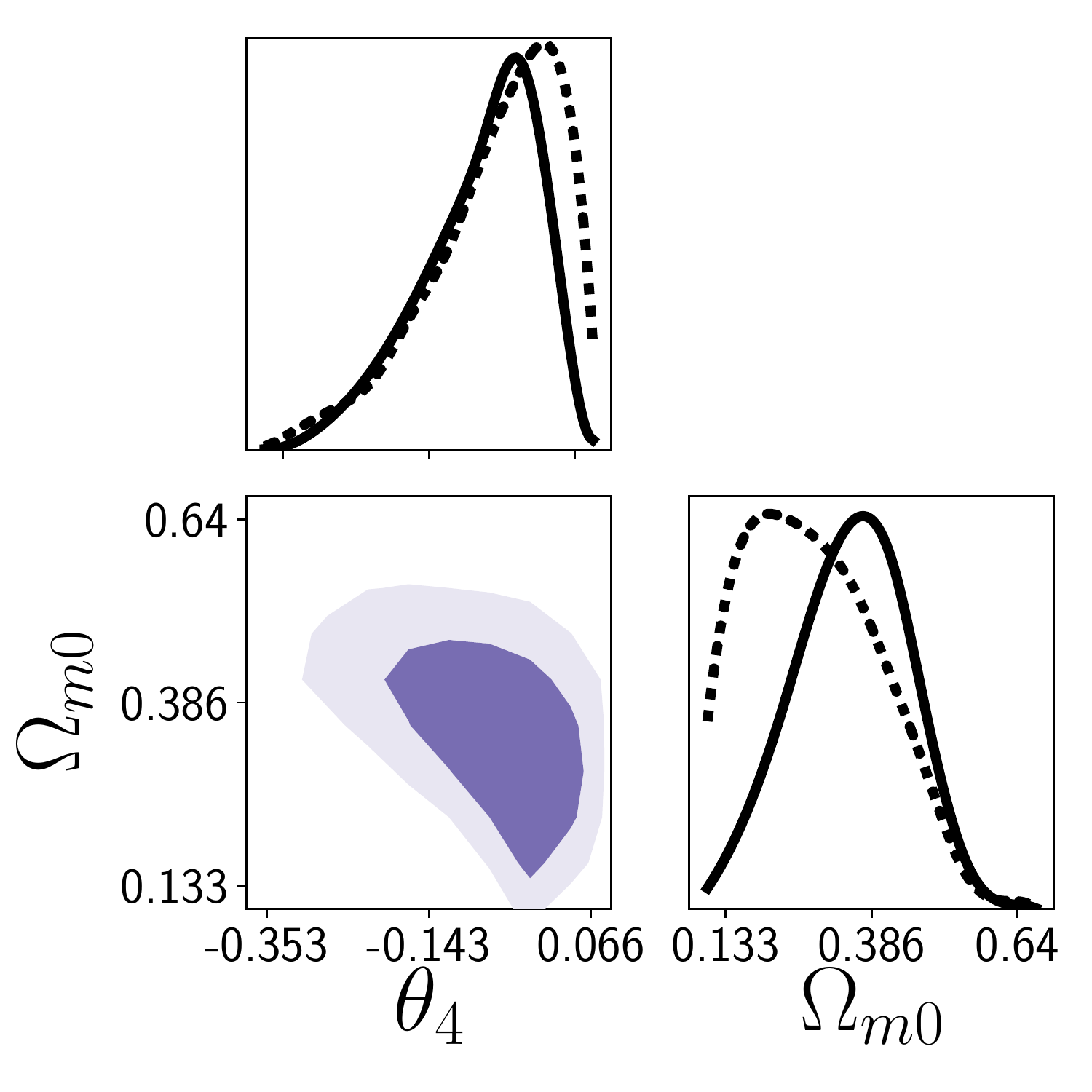}
  \caption{Marginalised posteriors for two of the parameters varied
    in the fit to JLA data: polynomial coefficient
    $\theta_4$ and matter density today $\Omn$ (obtained via \re{hbr}). Marginalised posteriors
    (2-dimensional contours and 1-dimensional solid lines) are
    biased by a volume effect. Maximised profiles (dashed lines) provide a reliable estimate.}
  \label{fig:vol_effect}
\end{figure}

\begin{table}[t]
  \centering
  \begin{tabular}{ccc}
    \hline
    \hline

    & $\theta_{4 }$ & $\Omn$ \\
    \hline

    Marginalised
    & $-0.07 ^{+0.11}_{-0.06}$
    & $ 0.35 ^{+0.11}_{-0.10}$
    \\
    Maximised
    & $0.03 ^{+0.04}_{-0.14}$
    & $0.24 ^{+0.14}_{-0.10}$ \\
    \hline
  \end{tabular}
  \caption{Marginalised and maximised statistics (68\% C.I.)
    associated to figure~\ref{fig:vol_effect}.}
  \label{tab:vol_effect}
\end{table}

Figure~\ref{fig:vol_effect} shows the marginalised posteriors for two
of the parameters varied in the SN data fit when determining
$\Omn$ from \re{hbr} by demanding that $h(z)$ is finite and real. The
1-dimensional plots show both the marginalised posteriors and the
maximised profile likelihood. Table~\ref{tab:vol_effect} shows the
mean together with the 68\% marginalised minimum credible intervals
\cite{Hamann:2007pi} (obtained by projecting all the points of the
multi-dimensional parameter space onto 1-dimensional histograms) as
well as the best fit together with the 68\% confidence intervals
(obtained by computing the maximum likelihood in each histogram bin).

Maximised confidence intervals describe the likelihood and are
interpreted from a frequentist point of view, while the marginalised
credible intervals are based on Bayesian posterior analysis. Maximised
profiles have the advantage of preserving the true peak of the
original multi-dimensional posterior probability. Marginalisation
instead favours regions of parameter space that have large volume in
the marginalised directions, and may lead to a misleading volume
effect \cite{Hamann:2007pi}. For example, consider the
$\theta_4$ - $\Omn$ plot in figure~\ref{fig:vol_effect}.  It is
instructive to compare the 2-dimensional posterior (keeping in mind
it is obtained by projecting from a higher-dimensional space) to the
1-dimensional profiles for $\Omn$.  While the global best fit is
$\Omn=0.24$ (given by the maximised profile, dashed line), there is a
large amount of volume at higher $\Omn$ values due to the large spread
in the $\theta_4$ direction. This pushes the marginalised (projected)
posterior mean to the larger value $\Omn=0.35$.

While the tension between the maximised and marginalised profiles is
not severe when the error bars are taken into account, as we see from
table~\ref{tab:vol_effect}, the differences are amplified in derived
parameters such as $\Omn$, and hence in $k_H(z)$ and $k_S(z_\ml,z_\ms)$.

\bibliographystyle{JHEPM}

\bibliography{refs}

\end{document}